\documentclass[12pt,a4paper,final]{iopart}

\usepackage{iopams}  
\usepackage{graphicx}
\usepackage{bm}
\usepackage{hyperref}

\expandafter\let\csname equation*\endcsname\relax
\expandafter\let\csname endequation*\endcsname\relax
\usepackage{amsmath}
\usepackage{float}

\usepackage{multirow}

\usepackage{tikz}

\usepackage{caption}
\usepackage{subcaption}
\usepackage[ruled]{algorithm2e}

\tikzstyle{line} = [draw, -latex']
\tikzstyle{cloud} = [draw, fill=red!20, node distance=3cm,minimum height=2em]
\tikzstyle{ball} = [circle,inner sep=5pt,draw]

\begin{document}

\title[\texttt{Gmunu}: Multigrid methods for solving Einstein field equations]{\texttt{Gmunu}: Toward multigrid based Einstein field equations solver for general-relativistic hydrodynamics simulations}

\author{Patrick Chi-Kit Cheong}
\ead{chi-kit.cheong@ligo.org}

\author{Lap-Ming Lin}
\ead{lmlin@cuhk.edu.hk}

\author{Tjonnie Guang Feng Li}
\ead{\mailto{tgfli@cuhk.edu.hk}}

\address{Department of Physics, The Chinese University of Hong Kong, Shatin, N. T., Hong Kong}

\begin{abstract}

We present a {new open-source axisymmetric} general relativistic hydrodynamics code \texttt{Gmunu} (\texttt{G}eneral-relativistic \texttt{mu}ltigrid \texttt{nu}merical solver) which uses a multigrid method to solve the elliptic metric equations in the conformally flat condition (CFC) approximation {on a spherical grid}.
Most of the existing relativistic hydrodynamics codes are based on formulations which rely on a free-evolution approach of numerical relativity, where the metric variables are determined by hyperbolic equations without enforcing the constraint equations in the evolution. 
On the other hand, although a fully constrained-evolution formulation is theoretical more appealing and should lead to more stable and accurate simulations, such an approach is not widely used because solving the elliptic-type constraint equations during the evolution is in general more computationally expensive than hyperbolic free-evolution schemes. 
Multigrid methods solve differential equations with a hierarchy of discretizations and its computational cost is generally lower than other methods such as direct methods, relaxation methods, successive over-relaxation. With multigrid acceleration, one can solve the metric equations on a comparable time scale as solving the hydrodynamics equations. This would potentially make a fully constrained-evolution formulation more affordable in numerical relativity simulations.
As a first step to assess the performance and robustness of multigrid methods in relativistic simulations, we develop a hydrodynamics code that makes use of standard finite-volume methods coupled with a multigrid metric solver to solve the Einstein equations in the CFC approximation. 
In this paper, we present the methodology and implementation of our code \texttt{Gmunu} and its properties and performance in some benchmarking relativistic hydrodynamics problems.

\end{abstract}

\vspace{2pc}
\noindent{\it Keywords}: Article preparation, IOP journals
\submitto{\CQG}

\section{Introduction}
In the past decade, numerical relativity has matured to the state that stable and robust numerical calculations of the Einstein equations with or without matter has become feasible. 
The spacetimes of many interesting astrophysical systems such as stellar core collapses and binary systems of compact objects have been accurately modeled (see, e.g., \cite{XNS2,2018MNRAS.480.5272C,2019arXiv190811258C} for recent reviews). 
In the standard $3+1$ decomposition of spacetime, the Einstein equations are split into a set of evolution equations and constraint equations. 
Nevertheless, one still has the freedom to choose the basic variables to evolve and reformulate the resulting systems of differential equations in order to improve the stability and accuracy of numerical simulations. 
This results in different formulations of numerical relativity, such as the so-called BSSN \cite{1995PhRvD..52.5428S, 1999PhRvD..59b4007B}, CCZ4 \cite{2003PhRvD..67j4005B}, and Z4c \cite{2010PhRvD..81h4003B} schemes, which are popular choices for numerical modelings. 
The practical applications of these different formulations are based on a free-evolution approach where the constraint equations are first solved for preparing the initial data and used subsequently only as an indicator to monitor the numerical accuracy during the evolution (see, e.g., \cite{2014PhRvD..89h4043M}).

Alternatively, one can also formulate the Einstein equations based on a fully constrained-evolution approach where the constraint equations are solved and fulfilled to within the discretization errors during the evolution. 
Despite the fact that a constrained-evolution approach is theoretical appealing, such an approach is not popular among numerical relativists since solving the elliptic-type constraint equations during the evolution is generally computational expensive. 
In contrast to the active development and applications of free-evolution formulations, the last proposed constrained formulation of the Einstein equations was already 15 years ago due to Bonazzola {\it et al.} \cite{2004PhRvD..70j4007B}. 
The fully constrained-evolution formulation of Bonazzola {\it et al.} \cite{2004PhRvD..70j4007B} has been employed to simulate pure gravitational wave spacetime \cite{2004PhRvD..70j4007B}, and also an oscillating neutron star by ignoring the back-reaction of the gravitational waves into the fluid dynamics \cite{2012PhRvD..85d4023C}. 
However, the application of this constrained scheme and assessment of its performance in modelling more generic dynamical spacetimes without symmetry is still a largely unexplored area.

It is worth to point out that the fully constrained scheme of Bonazzola {\it et al.} \cite{2004PhRvD..70j4007B} automatically reduces to the so-called conformally flat condition (CFC) approximation to general relativity \cite{wilson_mathews_2003,2008IJMPD..17..265I} if a tensor field $h_{ij}$ introduced in their formulation is set to zero (see \cite{XCFC} for a detailed discussion). 
The CFC approximation results in a simpler set of elliptic equations for the metric sector. 
Numerical simulations based on the CFC scheme have been successfully carried out for various astrophysical problems \cite{2002A&A...393..523D, 2002PhRvD..65j3005O, 2004ApJ...615..866S, 2012PhRvD..86f3001B, 2013ApJ...773...78B, 2014PhRvD..90b3002B, 2015MNRAS.453..287M} 
and the scheme has also been shown to be a good approximation to full general relativity in rotating iron core collapses \cite{2007CQGra..24S.139O}. 
However, the original CFC scheme suffers from mathematical non-uniqueness problems when the system is too compact.
In order to overcome the non-uniqueness issue, the scheme was reformulated and extended to the so-called extended CFC (xCFC) scheme so that the modelling of extreme spactimes such as black hole formation becomes possible \cite{XCFC,2019MNRAS.484.3307M,2012PhRvD..85d4023C}.

We have in our mind a motivation to experiment and develop our own general relativistic hydrodynamics code based on the fully constrained formulation of Bonazzola {\it et al.} \cite{2004PhRvD..70j4007B} (or other similar constrained formulations if available in the future) which maximizes the use of elliptic-type equations for the metric sector of the system in the evolution. 
In this paper, we take a first step along this direction by developing a relativistic hydrodynamics code based on the xCFC scheme. 
Although it is not fully general relativistic, the xCFC scheme contains a set of similar, but simpler, elliptic equations as the fully constrained formulation. 
We can thus use the xCFC scheme to evaluate the performance and robustness of our metric solver.

As already pointed out, it is known in general that solving the elliptic equations frequently in a fully constrained or xCFC scheme during a simulation is computationally expensive. 
Many numerical methods have been explored to deal with such elliptic systems, including finite-difference methods, different types of iterative solvers, and spectral methods (see \cite{CoCoA,CoCoNuT,XECHO} and references therein). 
A seminal work is due to Dimmelmeier {\it et al.} \cite{CoCoNuT} which combines a finite-difference grid and a spectral grid, on which the hyperbolic hydrodynamics and elliptic metric equations are solved, respectively.
However, even though the spectral method is known to be extremely fast and accurate, the metric solver is still one of the bottlenecks to slow down a hydrodynamics simulation as the communication between variables defined on the two different grids is time consuming especially in the multidimensional cases \cite{2006MNRAS.368.1609D,CoCoNuT}.
{In this work, we demonstrate that our nonlinear cell-centred multigrid method is not only an efficient strategy to solve the elliptic metric equations, but also can be straightforwardly used in hydrodynamical simulations. }

Multigrid methods solve differential equations with a hierarchy of discretizations and its computational cost is generally lower than other methods such as direct methods, relaxation methods, and successive over-relaxation \cite{MGbook}.
The multigrid strategy has been employed in a wide range of problems and it has also been used to generate initial data in numerical relativity \cite{2006gr.qc.....4100H,2012PhRvD..86j4053E,2014PhRvD..90h4043M}.
However, multigrid methods have not been applied in any constrained-evolution schemes for numerical relativity.
In order to couple to the matter directly, nonlinear cell-centred multigrid (CCMG) and the corresponding boundary treatments are needed, the latter of which is more complicated than the vertex-centred multigrid and is still being actively studied in the computational physics and applied mathematics.

Our aim is to construct a direct, rapid, and robust multidimensional metric solver which can be easily coupled to the matter based on the non-linear cell-centred multigrid strategy.
In this paper, we present the methodology and implementation of our {new open-source axisymmetric} general relativistic hydrodynamics code \texttt{Gmunu} (General-relativistic MUltigrid NUmerical solver), which solves the hydrodynamics equations using standard finite-volume methods and the xCFC metric equations using a multigrid approach {on a spherical grid}. 
{ \texttt{Gmunu} is written in the \texttt{Fortran90} programming language and is released as open source.}
To the best of our knowledge, this is the first relativistic hydrodynamics code that makes use of a multigrid solver in dynamical simulations. 
We also perform various benchmarking tests in relativistic hydrodynamics to assess the performance and robustness of our code.

The paper is organised as follows.
In \sref{sec:formulations} we outline the formalism we used in this work.
The details of the numerical settings and, the methodology, implementation of our hydrodynamics solver and our multigrid solver are presented in \sref{sec:numerical_methods} and \sref{sec:mg_solver} respectively.
The code tests and results are presented in { \sref{sec:hydro_test} and \sref{sec:spacetime}}.
{The performance of our multigrid solver is presented in \sref{sec:performance}.}
This paper ends with a discussion section in \sref{sec:conclusion}.

\section{\label{sec:formulations}Formulations}
\subsection{\label{sec:cfc_equations}Metric equations and Conformal flatness approximation}
We use the standard ADM 3+1 formalism \cite{2007gr.qc.....3035G,2008itnr.book.....A}. 
The metric can be written as
\begin{equation}
	ds^2 = -\alpha^2 dt^2 + \gamma_{ij} \left( dx^i + \beta^i dt \right)\left( dx^j + \beta^j dt \right),
\end{equation}
where $\alpha$ is the lapse function, 
$\beta^i$ is the spacelike shift vector 
 and $\gamma_{ij}$ is the spatial metric.
In the 3+1 formalism, the Einstein equations are split into a set of constraint equations which must be satisfied on every hypersurface
\begin{align}
	&R + K^2 - K_{ij}K^{ij} = 16 \pi E ,\\
	&\nabla_i(K^{ij} - \gamma^{ij} K) = 8 \pi S^i, 
\end{align}
and a set of the evolution equations for $\gamma_{ij}$ and the extrinsic curvature $K_{ij}$
\begin{align}
	\partial_t \gamma_{ij} = & - 2 \alpha K_{ij} + \nabla_i \beta_j + \nabla_j \beta_i  ,\\
	\partial_t K_{ij} = & - \nabla_i \nabla_j \alpha + \alpha \left( R_{ij} + KK_{ij} - 2 K_{ik}K^{k}_j \right) \\ \nonumber
	& + \beta^k\nabla_k K_{ij} + K_{ik}\nabla_j\beta^k + K_{jk} \nabla_i\beta^k \\ \nonumber
	& - 4\pi \alpha \left( 2S_{ij} - \gamma_{ij}\left( S^k_k - E \right) \right),
\end{align}
where $\nabla_i$ is the covariant derivative with respect to the three-metric $\gamma_{ij}$, $R_{ij}$ is the corresponding Ricci tensor, $R$ is the scalar curvature and $K$ is the trace of the extrinsic curvature $K_{ij}$.
For the matter sources, $E:= n_\mu n_\nu T^{\mu\nu}$, $S^{i}:= - n_\mu \gamma^i_\nu T^{\mu\nu}$ and $S^{ij}:= \gamma^i_\mu \gamma^j_\nu T^{\mu\nu}$, where $T^{\mu\nu}$ is the energy-momentum tensor
and $n_\mu$ is the unit normal vector of a spacelike hypersurface.

It is difficult to maintain the constraint equations in the numerical evolution of the evolution equations above because these ADM equations are numerically unstable.
There are serveral different re-formulations of $3+1$ numerical relativity that can lead to stable evolutions \cite{1995PhRvD..52.5428S, 1999PhRvD..59b4007B, 2003PhRvD..67j4005B, 2010PhRvD..81h4003B}.
However, these schemes are based on a free-evolution approach where the Einstein equations are evolved with hyperbolic-type equations. 
The constraint equations are only used for solving the initial data and serve as a monitor for numerical errors during the simulations. 
On the other hand, a fully-constrained evolution approach where the constraints are enforced at each time step is generally not favored as solving the elliptic-type constraint equations is computational expensive comparing to hyperbolic equations \cite{Press:1992:NRF:141273}.
We have a motivation to develop and experiment efficient multigrid solvers for elliptic-type metric equations that one needs in order to carry out fully-constrained evolutions for numerical relativity. 
As a first step towards this goal, our relativistic hydrodynamics code employs the xCFC scheme which is an improved version of the CFC approximation to general relativity.

In a CFC approximation \cite{CoCoA, XECHO}, the three metric $\gamma_{ij}$ is assumed to be decomposed according to   
\begin{equation}
	\gamma_{ij} := \psi^4 f_{ij}, 
\end{equation}
where $f_{ij}$ is a time independent flat background metric and $\psi$ is the conformal factor which is a function of space and time.
Another assumption is the maximal slicing condition of foliations
\begin{equation}
	K = 0.
\end{equation}
With these conditions, one can derive the time derivative of the conformal factor $\psi$ and also the extrinsic curvature $K_{ij}$
\begin{align}
	\partial_t \psi &= \frac{\psi}{6}\nabla_k\beta^k ,\\
	K_{ij} &= \frac{1}{2\alpha}\left( \nabla_i \beta_j + \nabla_j \beta_i - \frac{2}{3}\gamma_{ij}\nabla_k \beta^k \right).
\end{align}
The CFC approximation of the ADM equations can be reduced into five coupled non-linear elliptic equations
\begin{align}
	&\tilde{\Delta} \psi = \left( -2\pi {E} - \frac{1}{8}K^{ij}K_{ij} \right) \psi^{5} ,\\
	&\tilde{\Delta} (\alpha\psi) = \alpha\psi^5\left[ 2\pi \left( {E} + 2{S} \right) + \frac{7}{8}K_{ij}K^{ij}\right] ,\\
	&\tilde{\Delta} \beta^i + \frac{1}{3}\tilde{\nabla}^i \left( \tilde{\nabla}_j \beta^j \right) = 16 \pi \alpha \psi^{4} f^{ij} {S_i} + 2 \psi^{10} K^{ij}\tilde{\nabla}_j\left(\alpha \psi^{-6}\right),
\end{align}
where $\tilde{\nabla_i}$ and $\tilde{\Delta}$ are the covariant derivative and the Laplacian with respect to the flat three metric $f_{ij}$, respectively.  

The original CFC scheme suffers from mathematical non-uniqueness problems. 
The CFC scheme was later reformulated so that the elliptic equations are fully decoupled and the local uniqueness of the solution is guaranteed.
The reformulated CFC scheme is the so-called xCFC scheme \cite{XCFC}, which is the scheme that we implemented in \texttt{Gmunu}. 
In the xCFC scheme, one introduces a vector potential $X^i$, and the metric can be solved by the following equations:   
\begin{align}
	&\tilde{\Delta} X^i + \frac{1}{3}\tilde{\nabla}^i \left( \tilde{\nabla}_j X^j \right) = 8 \pi \tilde{S^i} \label{eq:X} ,\\
	&\tilde{\Delta} \psi = -2\pi \tilde{E}\psi^{-1} - \frac{1}{8}f_{ik}f_{jl}\tilde{A}^{kl}\tilde{A}^{ij} \psi^{-7}  \label{eq:psi} ,\\
	&\tilde{\Delta} (\alpha\psi) = (\alpha\psi)\left[ 2\pi \left( \tilde{E} + 2\tilde{S} \right) \psi^{-2} + \frac{7}{8}f_{ik}f_{jl}\tilde{A}^{kl}\tilde{A}^{ij} \psi^{-8} \right] \label{eq:alpha} ,\\
	&\tilde{\Delta} \beta^i + \frac{1}{3}\tilde{\nabla}^i \left( \tilde{\nabla}_j \beta^j \right) = 16 \pi \alpha \psi^{-6} f^{ij} \tilde{S_i} + 2 \tilde{A}^{ij}\tilde{\nabla}_j\left(\alpha \psi^{-6}\right) \label{eq:beta},
\end{align}
where $\tilde{E}:=\psi^6 E$, $\tilde{S_i}:=\psi^6 S_i$ and $\tilde{S}:=\psi^6 S$ are the rescaled fluid source terms.
The tensor field $\tilde{A}^{ij}$ can be approximated on the CFC approximation level by (see the Appendix of \cite{XCFC}): 
\begin{align}\label{eq:Aij}
	&\tilde{A}^{ij} \approx \tilde{\nabla}^i X^j + \tilde{\nabla}^j X^i - \frac{2}{3}\tilde{\nabla}_k X^k f^{ij}.
\end{align}

\subsection{General relativistic hydrodynamics equations}
The evolution equations for the matter are derived from the local conservations of the rest-mass and energy-momentum:
\begin{equation}
	\nabla_{\mu}(\rho u^{\mu}) = 0 \;\;\text{and}\;\; \nabla_\nu T^{\mu\nu} = 0,
\end{equation}
where $\rho$ is the rest-mass density of the fluid and $u^\mu$ is the fluid four-velocity. 
For a perfect fluid, the energy-momentum tensor is given by $T^{\mu\nu} = \rho h u^{\mu} u^{\nu} + P g^{\mu\nu}$, where $P$ is the pressure, $h = 1 + \epsilon + P/\rho$ is the enthalpy, and $\epsilon$ is the specific internal energy.
The three-velocity $v^i$ of the fluid as measured by the Eulerian observers of four-velocity $n^\mu$ is given by $v^i = \frac{u^i}{\alpha u^0} + \frac{\beta^i}{\alpha}$.
In the flux-conservative Valencia formulation (e.g., \cite{2000PhRvD..61d4011F}), the set of hydrodynamics equations are given by 
\begin{equation}
	\partial_t(\sqrt{\gamma} \bm{U}) + \partial_i(\sqrt{-g}\bm{F^i}) = \sqrt{-g}\bm{Q}, 
\end{equation}
where
\begin{align}
	&\bm{U} = \begin{bmatrix}
           D \\
        S_j \\
        \tau
        \end{bmatrix}
	= \begin{bmatrix}
           \rho W \\
        \rho h W^2 v_j \\
        \rho h W^2 - P - D
	\end{bmatrix}, \\
	&\bm{F^i} = \begin{bmatrix}
		D \left( v^i - \beta^i/\alpha \right) \\
		S_j \left( v^i - \beta^i/\alpha \right) + \delta^i_j P\\
		\tau \left( v^i - \beta^i/\alpha \right) + Pv^i
        \end{bmatrix}, \\
	&\bm{Q} = \begin{bmatrix} \label{eq:grav_source}
           0 \\
		T^{\mu\nu}\left( \frac{\partial g_{\nu j}}{\partial x^\mu} - \Gamma^{\lambda}_{\mu\nu} g_{\lambda j} \right) \\
		\alpha \left( T^{\mu 0} \frac{\partial \ln \alpha}{\partial x^\mu} - T^{\mu\nu} \Gamma^{0}_{\mu\nu} \right) \\
        \end{bmatrix}
	.
\end{align}
As shown in Eq.~\eqref{eq:grav_source}, the source terms $\bm{Q}$ contain the time derivatives of the metric quantities.
In order to reduce the accumulated error due to the time update, it is good to avoid the time derivatives in the code.
We can rewrite the $Q_j$ terms into compact form \cite{CoCoNuT}
\begin{equation}
        Q_j = \frac{1}{2}T^{\mu\nu}\frac{\partial g_{\mu \nu}}{\partial x^j} .
\end{equation}
It is also possible to bypass the time derivatives in the $Q_\tau$ term (i.e., the last element of the vector in Eq.~\eqref{eq:grav_source}):
\begin{align}
        Q_\tau = &T^{00} \left( K_{ij} \beta^i \beta^j - \beta^k \partial_k \alpha \right) \\ \nonumber
        &+ T^{0j} \left( 2 K_{jk} \beta^k - \partial_j \alpha \right) + T^{ij}K_{ij}.
\end{align}

In order to adapt to the extended CFC scheme, we have to evolve the conformal transformed conserved quantities, the (non-conformal transformed) conserved variables and thus the primitive variables will be updated once the conformal factor $\psi$ is solved.

In particular, we define $\bm{\tilde{U}} \equiv \psi^6 \bm{U}$, $\bm{\tilde{F}^i} \equiv \psi^6 \bm{F^i}$ and $\bm{\tilde{Q}} \equiv \psi^6 \bm{Q}$.
We can then reformulate the hydrodynamics equations as
\begin{equation}
        \frac{\partial \bm{\tilde{U}}}{\partial t} + \frac{1}{r^2}\frac{\partial}{\partial r}\left(\alpha r^2 \bm{\tilde{F}^r}\right) + \frac{1}{\sin \theta}\frac{\partial}{\partial \theta}\left(\alpha \sin \theta \bm{\tilde{F}^\theta}\right) = \alpha\bm{\tilde{Q}} .
\end{equation}
In the case that we need to solve the metric, we pass the conformal conserved quantities into the metric solver.
Both metric quantities and primitive hydrodynamic variables will then be updated.

\section{\label{sec:numerical_methods}Numerical methods and implementation}
We use spherical polar coordinates \{$r,\theta,\phi$\} and adopt the axial symmetry with cell-centered discretization.  
In particular, the coordinate grid covers $0<r<r_{\text{max}}$ and $0<\theta<\pi/2$ and we discretize it into $n_r \times n_\theta$ cells with uniform coordinate grid spacing, i.e. $\Delta r = r_{\text{max}} / n_r$ and $\Delta \theta = \pi /2 n_\theta$. 

\texttt{Gmunu} solves the general relativistic hydrodynamic equations by using standard high-resolution shock-capturing (HRSC) schemes \cite{1991PhRvD..43.3794M}. 
In particular, various different cell-interface reconstruction methods and Riemann solvers are implemented and tested.   
For the cell-interface reconstruction, we have implemented piecewise constant scheme (PC), monotonized-central limiter (MC), $5^\text{th}$ order weighted-essentially nonoscillatory scheme (WENO5) \cite{2009SIAMR..51...82S} and $5^\text{th}$ order monotonicity preserving scheme (MP5) \cite{1997JCoPh.136...83S}.
For the Riemann solver, the Rusanov flux (also known as Total variation diminishing Lax-Friedrichs scheme (TVDLF)) \cite{1996ApL&C..34..245T}, Harten-Lax-van Leer (HLL) \cite{Harten1997}, Harten-Lax-van Leer-Einfeldt (HLLE) \cite{HLLE1, HLLE2}, Marquina flux formula \cite{1996JCoPh.125...42D} have been implemented.
For the recovery of the primitive variables $(\rho, v_i, P)$ from the conservative variables $(D, S_i, \tau)$, we follow the formulation presented in the Appendix C in \cite{2013PhRvD..88f4009G} and use Regula-Falsi method to find the root.
This formulation was shown to be not only robust, accurate and efficient, but also suitable for different kinds of relativistic hydrodynamical simulations \cite{2013rehy.book.....R,2013PhRvD..88f4009G}.
For the region outside the star, we fill with an ``atmosphere'' of density $\rho_{\text{atmo}} = 10^{-6} \rho_\text{max}(t=0)$, where $\rho_\text{max}(t=0)$ is the maximum density over the whole computational domain initially at time $t=0$, and use the standard atmosphere treatment during the simulation.
In particular, if the density at a particular grid drops below $\rho_{\rm atmo}$, the density at that grid is reset to $\rho_{\rm atmo}$. 
The velocity of that grid is also set to be zero and other primitive variables such as the specific internal energy $\epsilon$ is updated accordingly.

{Unless otherwise specified, }the simulations reported in this paper were preformed with HLLE Riemann solver and the MC reconstruction method. 
We also use a $3^\text{rd}$ order Runge-Kutta integrator for the time integration. 
It is not necessary to solve the metric at each time step \cite{CoCoA, XECHO}. 
To speed up the simulations, we solve the metric at every 50 time steps, and extrapolate the metric in between. 
{Detailed analysis of frequency of solving the metric can be found in \sref{sec:compare}.}
A practical difficulty related to spherical coordinates is that the convergence of grid points near the pole axis puts a severe constraint on the time step imposed by the Courant-Friedrichs-Levy stability condition in numerical simulations. 
In order to increase the size of the time steps in our simulations, we treat the first 10 grid points, which cover about $5\%$ of the stellar radius, as a spherically symmetric core (i.e., only radial motions are allowed).
This should be a good approximation as non-radial fluid motions in the core is negligibly small.
We used the open-source code \texttt{XNS} \cite{XECHO,XNS1,XNS2,XNS3}, which is also based on the CFC approximation, to generate initial neutron-star models for our dynamical simulations.

\section{\label{sec:mg_solver}Nonlinear cell-centered multigrid metric solver}
\subsection{\label{sec:metric_solver}Metric solver with xCFC scheme}
By following \cite{XCFC}, the metric equations Eqs.~\eqref{eq:X}-\eqref{eq:beta} can be decoupled, thus they can be solved in a hierarchical way and the local uniqueness is guaranteed.
Once the time integration at each time step for the hydrodynamics equations is completed, the conformally rescaled hydrodynamical conserved variables $(\tilde{D}, \tilde{S}_i, \tilde{\tau})$ are updated, and they will be used to solve the metric equations.
The steps for solving the metric equations are summarized in the following:
\begin{enumerate}
	\item Solve Eq.~\eqref{eq:X} for the vector potential $X^i$ from the conserved variables $\tilde{S}_i$.
	\item Calculate the tensor $\tilde{A}^{ij}$ in Eq.~\eqref{eq:Aij} from the vector potential $X^i$.
	\item Solve Eq.~\eqref{eq:psi} for the conformal factor $\psi$.
	\item With the updated conformal factor $\psi$, calculate the conserved variables $({D}, {S}_i, {\tau})$ and thus convert the conserved variables to the primitive variables $({\rho}, {v}_i, {P})$.
		Then $\tilde{S}$ can be worked out consistently.
	\item Solve Eq.~\eqref{eq:alpha} for the lapse function $\alpha$.
	\item Solve Eq.~\eqref{eq:beta} for the shift vector $\beta^i$.
\end{enumerate}

To solve the metric equations, appropriate boundary conditions at the origin and outer computational boundary are required. 
{In the practical simulations of spheric-like astrophysical systems (e.g. isolated neutron star and core-collapse supernova), as the spacetime at the star surface is close to spherically symmetric, setting the Schwarzschild solution as the outer boundary condition is usually a good approximation \cite{CoCoA, CoCoNuT}.}
In order to let the solution fall off as the Schwarzschild solution at large distance, we require $\psi = \frac{C}{r} + 1$, $\alpha = \frac{C}{r} + 1$ and $X^i = \beta^i = 0$ at the outer boundary ($r = r_{\text{max}}$). In \texttt{Gmunu}, we impose the boundary conditions
\begin{align}
	\frac{\partial \psi}{\partial r} \Big|_{r_{\text{max}}} &= \frac{1-\psi}{r}, \\
	\frac{\partial \alpha}{\partial r} \Big|_{r_{\text{max}}} &= \frac{1-\alpha}{r}, \\
	\beta^i \Big|_{r_{\text{max}}} &= 0, \\
	X^i \Big|_{r_{\text{max}}} &= 0 .
\end{align}
More detailed implementation of the metric equations and boundary conditions at the axis and the origin can be found in \ref{appendix:metric}.

\subsection{An overview of multigrid}
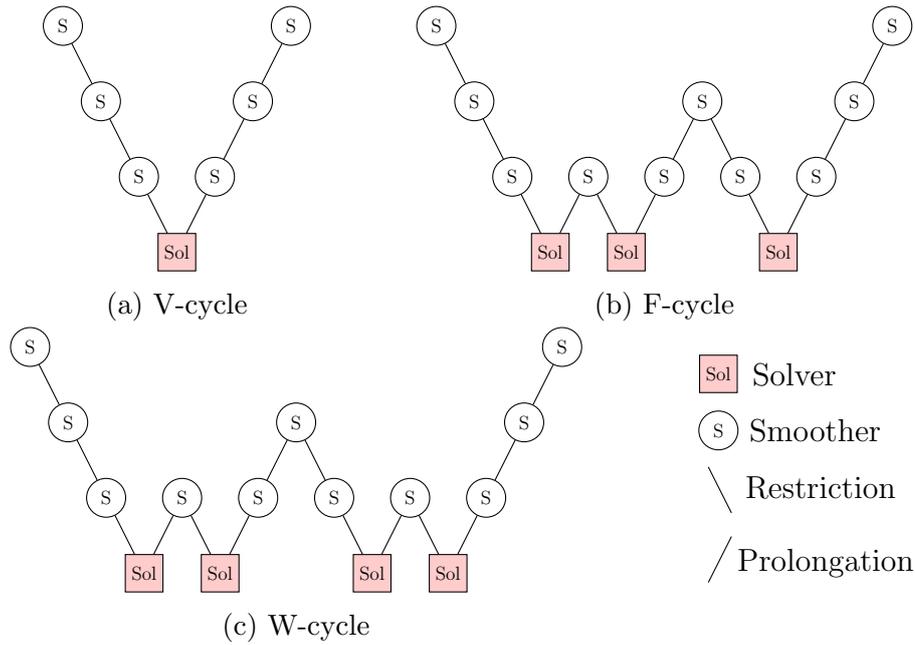
\begin{figure}[h]
	\centering
	\begin{subfigure}{0.3\columnwidth}
		\centering
		\begin{tikzpicture}[scale=0.5, every node/.style={scale=0.6}]
			\node [ball] (n7) at (3,6) {S};
			\node [ball] (n6) at (2,4) {S};
			\node [ball] (n5) at (1,2)  {S};
			\node [cloud] (n4) at (0,0)  {Sol};
			\node [ball] (n3) at (-1,2) {S};
			\node [ball] (n2) at (-2,4)  {S};
			\node [ball] (n1) at (-3,6)  {S};

			\foreach \from/\to in {n1/n2,n2/n3,n3/n4,n4/n5,n5/n6,n6/n7}
			\draw (\from) -- (\to);
		\end{tikzpicture}
		\caption{V-cycle}
	\end{subfigure}
	\begin{subfigure}{0.5\columnwidth}
		\centering
		\begin{tikzpicture}[scale=0.5, every node/.style={scale=0.6}]
			\node [ball] (n13) at (9,6) {S};
			\node [ball] (n12) at (8,4) {S};
			\node [ball] (n11) at (7,2) {S};
			\node [cloud] (n10) at (6,0)  {Sol};
			\node [ball] (n9) at (5,2) {S};
			\node [ball] (n8) at (4,4) {S};
			\node [ball] (n7) at (3,2)  {S};
			\node [cloud] (n6) at (2,0)  {Sol};
			\node [ball] (n5) at (1,2) {S};
			\node [cloud] (n4) at (0,0)  {Sol};
			\node [ball] (n3) at (-1,2) {S};
			\node [ball] (n2) at (-2,4)  {S};
			\node [ball] (n1) at (-3,6)  {S};
		
			\foreach \from/\to in {n1/n2,n2/n3,n3/n4,n4/n5,n5/n6,n6/n7,n7/n8,n8/n9,n9/n10,n10/n11,n11/n12,n12/n13}
			\draw (\from) -- (\to);
		
		\end{tikzpicture}
			\caption{F-cycle}
	\end{subfigure}
	
	\begin{subfigure}{0.6\columnwidth}
		\centering
		\begin{tikzpicture}[scale=0.5, every node/.style={scale=0.6}]
			\node [ball] (n15) at (11,6) {S};
			\node [ball] (n14) at (10,4) {S};
			\node [ball] (n13) at (9,2)  {S};
			\node [cloud] (n12) at (8,0)  {Sol};
			\node [ball] (n11) at (7,2) {S};
			\node [cloud] (n10) at (6,0)  {Sol};
			\node [ball] (n9) at (5,2) {S};
			\node [ball] (n8) at (4,4) {S};
			\node [ball] (n7) at (3,2)  {S};
			\node [cloud] (n6) at (2,0)  {Sol};
			\node [ball] (n5) at (1,2) {S};
			\node [cloud] (n4) at (0,0)  {Sol};
			\node [ball] (n3) at (-1,2) {S};
			\node [ball] (n2) at (-2,4)  {S};
			\node [ball] (n1) at (-3,6)  {S};
		
			\foreach \from/\to in {n1/n2,n2/n3,n3/n4,n4/n5,n5/n6,n6/n7,n7/n8,n8/n9,n9/n10,n10/n11,n11/n12,n12/n13,n13/n14,n14/n15}
			\draw (\from) -- (\to);
		
		\end{tikzpicture}
			\caption{W-cycle}
	\end{subfigure}
	\begin{subfigure}{0.3\columnwidth}
		\begin{tikzpicture}[scale=0.5]
			\node [cloud,scale=0.6] (Sol) at (0,3)  {Sol};
			\node [ball,scale=0.6] (S) at (0,1.5)  {S};

			\node [ball,scale=0.6,opacity=0] (A) at (-0.5,1)  {S};
			\node [ball,scale=0.6,opacity=0] (AB) at (0,0)  {S};
			\node [ball,scale=0.4,opacity=0] (B) at (0.5,-1)  {S};
			\node [ball,scale=0.6,opacity=0] (BC) at (0,-2)  {S};
			\node [ball,scale=0.6,opacity=0] (C) at (-0.5,-3)  {S};

			\foreach \from/\to in {A/B,B/C}
			\draw (\from) -- (\to);

			\node [right of=S] (cap1) {\ \ \ \ Smoother};
			\node [right of=Sol] (cap2) {Solver};
			\node [right of=AB] (cap3) {\ \ \ \ \ Restriction};
			\node [right of=BC] (cap4) {\ \ \ \ \ \ Prolongation};
		
		\end{tikzpicture}
	\end{subfigure}

	\caption{ 
		Three types of (4-grid) cycles can be used in multigrid methods. 
		``S'' denotes smoothing while ``Sol'' denotes solving the equation directly.
		Each descending line $\setminus$ represents restriction and each ascending line / represents prolongation.
		The key ingredients of multigrid solver includes a cycle framework, restriction and prolongation operators, smoothers and solvers.
	}
	\label{fig:MG_cycles}	
\end{figure}

Multigrid is an efficient method for solving elliptic partial differential equations with low memory and work complexity.
Different modes are filtered out with different rates at different resolutions.
For some low-frequency modes, it is computationally expensive to compute directly on a high-resolution discretization.
However, it can be done efficiently at a low-resolution discretization.
The main concept of the multigrid method is to solve the problem recursively with a series of coarse grids.
It is noted that the multigrid method is not a single method but a strategy with many possible implementations. 
However, the elements needed to construct a multigrid solver is more or less the same for different implementations. 
For instance, as shown in \fref{fig:MG_cycles}, the key ingredients of multigrid solver includes
(1) {a \emph{cycle framework}, the backbone of the whole solver which describe the structure of the multigrid solver (see \fref{fig:MG_cycles} for examples),}
(2) inter grid transfer operators to connect the solver with different levels, where the operators that map the values from a fine to a coarse grid are called \emph{restriction} and the mapping from a coarse to the fine grid are called \emph{prolongation}.
(3) \emph{smoothers} to smooth the solutions at different resolutions and
(4) \emph{direct solvers} to obtain the solutions at the coarsest level.

\subsection{Nonlinear multigrid: The Full Approximation Scheme}
Due to the nonlinearity of the Einstein equations needed to be solved (i.e. Eqs.~\eqref{eq:X}-\eqref{eq:beta}), nonlinear multigrid method is required.
The implementations for the multigrid for non-linear elliptic equations are different from the linear cases.
Two well-known methods for solving non-linear partial differential equations with multigrid techniques are the Full Approximation Scheme (FAS) \cite{MR431719} and Newton-multigrid (Newton-MG) \cite{briggs2000multigrid,trottenberg2001multigrid}.
The two methods are widely used and obtained successes in various problems.
We refer the interested reader to \cite{BRABAZON20141619} for a detailed comparison of the two methods.
Due to the fact that the memory used is low in FAS, the current version of \texttt{Gmunu} adopts the FAS algorithm (see \cite{MR431719,BRABAZON20141619,press1996numerical} and references therein for more details).

{Here we briefly outline how the Full Approximation Scheme works.
To solve an nonlinear elliptic equation $\mathcal{L}(u) = f$, where $\mathcal{L}$ is some elliptic operator, $u$ is the solution and $f$ is the source term, we can discretize the equation on a grid with resolution $h$ as
\begin{equation}
 \mathcal{L}_h(u_h) = f_h.
\end{equation}
Suppose we have obtained an approximate solution $\tilde{u}_h$ through the smoothing processes (note that smoothers is a kind of solver, but slow), we can find the desired correction $e_h$ so that the equation $\mathcal{L}_h(\tilde{u}_h + e_h) = f_h$ is solved.
The residual $r_h$ is defined by
\begin{equation} \label{eq:res_h}
\begin{split}
r_h := &\mathcal{L}_h(\tilde{u}_h + e_h) - \mathcal{L}_h(\tilde{u}_h) \\
= &f_h - \mathcal{L}_h(\tilde{u}_h)
\end{split}
\end{equation}
By transfering the first line of the Eq. \eqref{eq:res_h} to a coarse grid with resolution $2h$ through the restriction operator $\mathcal{R}$, we have
\begin{equation} \label{eq:res_2h} 
 {\mathcal{L}_{2h}(u_{2h}) = \mathcal{R}(r_{h})- \mathcal{L}_{2h}(\mathcal{R}(\tilde{u}_h)).}
\end{equation}
Note that the RHS of Eq. \eqref{eq:res_2h} can now be treated as a new source term, that is, $ f_{2h} := \mathcal{R}(u_{h})- \mathcal{L}_{2h}(\mathcal{R}(\tilde{u}_h))$.
Let $v$ denote the approximate solution of Eq. \eqref{eq:res_2h}, we can then obtain the coarse-grid correction:
\begin{equation}
\tilde{e}_{2h} = v - \mathcal{R}(u_{h}),
\end{equation}
and thus the new approximate solution on resolution $h$ is
\begin{equation}
\begin{split}
\tilde{u}^{\text{new}}_{h} = & \tilde{u}_{h} + \mathcal{P}(\tilde{e}_{2h}) \\
= & \tilde{u}_{h} + \mathcal{P}(v - \mathcal{R}(u_{h})),
\end{split}
\end{equation}
where $\mathcal{P}$ is the prolongation operator.
Note that FAS can also be used to solve linear elliptic equations.
}
Algorithm \ref{alg:MG} shows the pseudocode of a single cycle of the non-linear multigrid elliptic partial differential equation solver implemented in \texttt{Gmunu}.

\begin{algorithm}[H]
	\SetKwFunction{MG}{MG}
	\SetKwFunction{Smoothing}{Smoothing}
	\SetKwFunction{Solve}{Solve}
	\SetKwFunction{Restriction}{Restriction}
	\SetKwFunction{Prolongation}{Prolongation}
	\tcc{ Before solving any equations, we have to initialize the $\gamma$ based on what cycle type we want to use. In particular:}
	\tcc{ V-cycle: set $\gamma = 1$  }
	\tcc{ W-cycle: set $\gamma = 2$  }
	\tcc{ F-cycle: set $\gamma = 2$  }
	\tcc{   }
	$\MG(u_{l},f_{l})$ \\
	\eIf{$l = 1$}{ 
		$u_l \leftarrow \Solve(u_l)$  \\
		\If{ F-cycle }{	$\gamma \leftarrow 1 $}
		}{
		
		$u_l \leftarrow \Smoothing(u_l,f_l)$ \tcc*[r]{Pre-smoothing}
		$r_l \leftarrow f_l - L_l(u_l)$ \tcc*[r]{calculating the residual}
		
		$r_{l-1} \leftarrow \Restriction(r_l)$ \tcc*[r]{ restriction of the residual}
		$u_{l-1} \leftarrow \Restriction(u_l)$ \tcc*[r]{ restriction of the solution}

		$f_{l-1} \leftarrow r_{l-1} + L_{l-1}(u_{l-1})$ \;
		$v \leftarrow u_{l-1}$ \;
		\tcc{ Recursive call for the coarse grid correction }
		\For{$i = 1$ \KwTo $\gamma$}{
			$v \leftarrow \MG(v,f_{l-1})$
		}
		
		$ u_{l} \leftarrow u_{l} + \Prolongation(v - u_{l-1})$ \tcc*[r]{ Prolongation }

		$u_l \leftarrow \Smoothing(u_l,f_l)$ \tcc*[r]{ Post-smoothing } 

		\If{ finest level and F-cycle}{
			$\gamma \leftarrow 2$}
	}
	\caption{ 
	A single cycle of the non-linear multigrid elliptic partial differential equation solver implemented in \texttt{Gmunu}.
	}
	\label{alg:MG}
\end{algorithm}

\subsection{Cell-centered discretization and intergrid transfer operators}
As mentioned in \sref{sec:cfc_equations}, the source terms of the metric equations consist of the hydrodynamical variables.
Meanwhile, since \texttt{Gmunu} solves the hydrodynamics with finite volume approach, the grids are discretized with cell-centred discretization.
On the other hand, the discretization for the metric solver is in general different from the hydrodynamics sector.
For instance, for the pseudospectral method, the choice of grid points has to be consistent with the basis functions which can not be chosen arbitrarily.
The corresponding interpolation or extrapolation are needed so that the hydrodynamical variables can be passed correctly into the metric solver (e.g. \cite{CoCoNuT, XECHO}), which might be another bottleneck of the computational time of the simulations.
In order to adapt the grid of the hydrodynamics sector so that the hydrodynamical variables can be passed into the metric solver without any interpolation or extrapolation, we implemented the cell-centred multigrid (CCMG) \cite{Mohr2004,MGbook}, which is one of the novelties of \texttt{Gmunu}.

Constructing a cell-centred multigrid solver is non-trivial.
Unlike the vertex-centred case, in which a node of the coarse grid is also a node of the fine grid, the nodes on coarser grids do not form a subset of the fine grid nodes in the case of cell-centred discretization.
The choices of inter-grid transfer operators and the boundary condition implementation are different from the vertex-centred cases.
There are also many possible approaches for different situations.
Indeed, constructing problem-independent efficient cell-centred transfer operators is still under an active research area in computational physics and applied mathematics (see \cite{Mohr2004} and references therein).

There are many possible choices of restriction and prolongation operators and they cannot be chosen arbitrarily \cite{Mohr2004,MGbook}.
In \texttt{Gmunu}, we adopt the most standard combination, i.e., piecewise constant restriction (\fref{fig:pc}) and bi-linear prolongation (\fref{fig:bilinear}).
\begin{figure}
	\centering
	\begin{subfigure}{0.4\columnwidth}
		\centering
\begin{equation*}
	\frac{1}{4} \left]\begin{array}{ccccc}
		  &   &   &   &  	\\
		  & 1 &   & 1 &  	\\
		  &   & * &   &  	\\
		  & 1 &   & 1 &  	\\
		  &   &   &   &  
	\end{array}\right[^h_{2h}
\end{equation*}
		\caption{piecewise constant restriction}
		\label{fig:pc}
	\end{subfigure}
	\begin{subfigure}{0.4\columnwidth}
		\centering
\begin{equation*}
	\frac{1}{16} \left]\begin{array}{ccccc}
		1 & 3 &   & 3 & 1	\\
		3 & 9 &   & 9 & 3	\\
		  &   & * &   &  	\\
		3 & 9 &   & 9 & 3	\\
		1 & 3 &   & 3 & 1
	\end{array}\right[^h_{2h}
\end{equation*}
		\caption{bi-linear prolongation}
		\label{fig:bilinear}
	\end{subfigure}

	\caption{ 
	The stencil notation of the interpolation operators implemented in \texttt{Gmunu}.
	The ``*'' denotes the location of the coarse grid node.
	The notation shows the weighting of the value which are the neighbors of the coarse grid node ``*''.
	}
	\label{fig:interpolation}	
\end{figure}

\subsection{Key features of the nonlinear cell-centered multigrid metric solver in \texttt{Gmunu}}
This section shortly summarizes the key features of the metric solver implemented in \texttt{Gmunu}.
In our multigrid metric solver, we adopt the Full Approximation Storage (FAS) to deal with the nonlinear metric equations with V-, W- and F-cycle implemented.
For the smoother and solvers, we use the standard red-black {nonlinear} Gauss-Seidel relaxation {\cite{press1996numerical}}. 
In particular, the smoother consists of 15-times relaxations and the direct solver consists of 200-times relaxations.
For the inter-grid transfer operators, we adopt piecewise constant restriction (\fref{fig:pc}) and bi-linear prolongation (\fref{fig:bilinear}).
Note that multigrid solvers are iterative solvers.
Practically, this function has to be called until the solution converges, i.e., when the $L_\infty$ or $L_1$ norm of the residual is below some chosen threshold value.

\section{\label{sec:hydro_test}Code tests 1: Hydrodynamics solver}
We present a set of numerical tests of \texttt{Gmunu}.
\Tref{tab:models} lists the models we used in various tests.
The name of the models are originally defined in the literature \cite{2006MNRAS.368.1609D,XCFC}.
All the models are constructed with the polytropic equation of state with $\Gamma=2$ and $K=100$ (in units of $(c=G=M_{\odot}=1)$).
\begin{table}[h]
  \begin{center}
    \caption{
	  The equilibrium neutron star models we used in this paper.
	  The name of the models are originally defined in the literature \cite{2006MNRAS.368.1609D,XCFC}.
	  ``BU'' represents a sequence of fixed central rest-mass density $\rho_c = 1.28 \times 10^{-3}$ uniformly rotating models, and ``SU'' is a nonrotating unstable model.
	  All the models are constructed with the polytropic equation of state with $\Gamma=2$ and $K=100$.
	  $\Omega$ is the angular velocity; $M$ is the gravitational mass; $r_e$ and $r_p$ are the equatorial and polar radii, respectively. 
	  Unless otherwise noted, we use units where $c=G=M_\odot=1$.
	  }
    \label{tab:models}
    \begin{tabular}{l|c|c|c|c|c} 
\br
	    Model & $\rho_c$ $[10^{-3}]$ & $\Omega$ $[10^{-2}]$ & $M$ $[M_{\odot}]$ & $r_{e}$ & $r_{p}/r_{e}$ \\
      \hline
	BU0 & $1.28$ & $0.000$ & $1.400$ & 8.13 & 1.00 \\
	SU & $8.00$ & $0.000$ & $1.447$ & 4.27 & 1.00  \\
	BU2 & $1.28$ & $1.509$ & $1.468$ & 8.56 & 0.90  \\
	BU8 & $1.28$ & $2.633$ & $1.693$ & 11.30 & 0.60  \\
\br
    \end{tabular}
  \end{center}
\end{table}

{We first perform two standard tests in a static background metric, namely, (1) the planar shocktube problem for special-relativistic hydrodynamics \cite{Martí2003} and (2) the evolution of a static stable neutron star on a fixed background metric in the Cowling approximation \cite{2004MNRAS.352.1089S}.
In the following, we show the performance of \texttt{Gmunu} in these tests.
}

\subsection{\label{sec:hydro_test_shocktube} Relativistic shocktube }
\begin{figure}[h]
	\centering
	\includegraphics[width=0.6\columnwidth, angle=0]{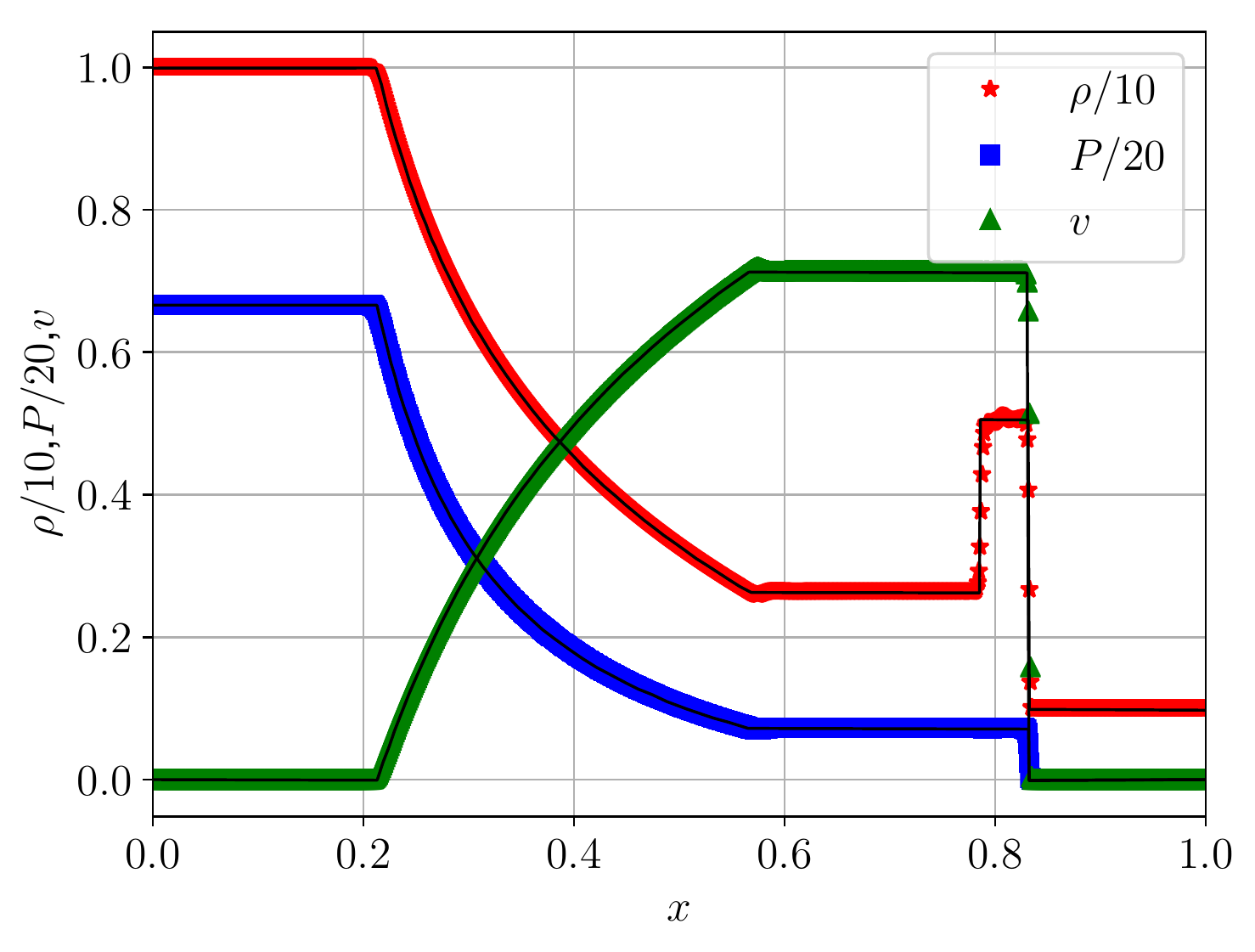}
	\caption{
		The density (red stars), pressure (blue squares) and velocity (green triangles) are shown at $t=0.4$ for the relativistic shocktube test problem.
		The solid lines are the analytic solutions.
		The numerical results obtained by \texttt{Gmunu} agree very well with the analytic solutions.
		}
	\label{fig:shocktube}	
\end{figure}

The shocktube problem is a standard test to access the shock-capturing ability of a relativistic hydrodynamics code.
In this test, we follow the setup proposed in \cite{Martí2003}.
In particular, we assume flat spacetime and perform the simulation with planar geometry, for $ x = [0,1] $ with 1000 grid points.
For the matter, we use the $\Gamma$-law equation of state with $\Gamma = 5/3$, and the initial conditions are set in \tref{tab:shocktube}. 
\begin{table}[h]
  \begin{center}
    \caption{Initial conditions for the relativistic shocktube problem.}
    \label{tab:shocktube}
    \begin{tabular}{l|c} 
\br
      $x < 0.5$ & $x > 0.5$\\
      \hline
      $\rho = 10$ & $\rho = 1$\\
      $P = 13.33$ & $P = 0$ \\
      $v = 0$ & $v = 0$ \\
\br
    \end{tabular}
  \end{center}
\end{table}
In this test, we used HLLE as Riemann solver with WENO5 reconstruction and RK3 for time integration.
\Fref{fig:shocktube} shows the comparison between the numerical results and the analytic solutions for the density, pressure and velocity profiles at $t = 0.4$.
The figure shows that our numerical results agree with the analytic solutions very well.

\subsection{\label{sec:hydro_test_Cowling} Spherical star evolutions with static metric}
The test we report here is the evolution of a stable spherically symmetric neutron star on a static background metric.
Although it is an equilibrium solution of the Einstein equations, the diffusion at the contact discontinuity of the neutron star surface triggers the natural oscillation modes which were well studied for this kind of polytropic star \cite{2004MNRAS.352.1089S}.
The oscillation modes of a neutron star can be obtained approximately by perturbation calculations in the Cowling approximation \cite{2002ApJ...568L..41Y,2004MNRAS.352.1089S,2005MNRAS.356..217Y}. 
They can also be extracted by nonlinear hydrodynamical simulation with a fixed background metric.
By comparing the simulated results with the oscillation modes obtained in the Cowling approximation, we can thus focus and check the correctness of our hydrodynamics solver on a fixed background spacetime.
In this test, we simulate the stable spherically symmetric neutron star BU0 model as mentioned in the \tref{tab:models}.
We setup a 1D run with the resolution $n_r \times n_{\theta} = 640 \times 1$ and keep the background metirc fixed during the simulation.

\begin{figure}[h]
	\centering
	\includegraphics[width=0.8\columnwidth, angle=0]{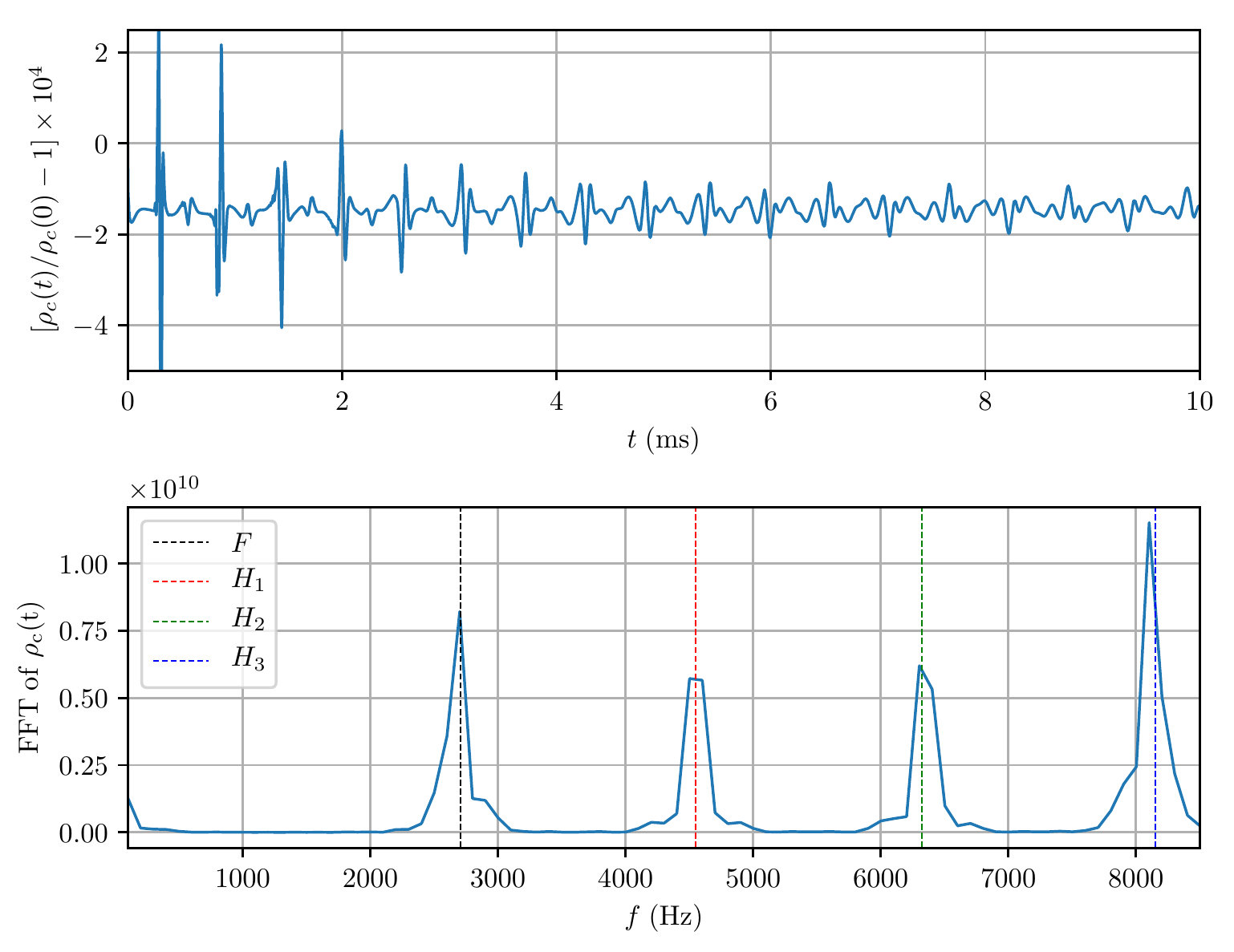}
	\caption{
		The evolution of a stable spherical symmetric neutron star (TOV star) with the resolution $n_r \times n_{\theta} = 640 \times 1$.
                \emph{Upper panel}: The relative variation of the central density in time.
                \emph{Lower panel}: The Fourier transform of the central density. The vertical lines represent the frequencies of the oscillation modes calculated in the Cowling approximation. 
	}
	\label{fig:Cowling_BU0}	
\end{figure}

The upper panel of \fref{fig:Cowling_BU0} shows the relative variation of the central density $\rho_c$ as a function of time.
The relative variation of the density is of the order $10^{-4}$.
The oscillation modes extracted from the simulation also agree very well with those obtained in the Cowling approximation.

\section{\label{sec:spacetime}Code tests 2: Metric solver}
{ After demonstrating that \texttt{Gmunu} can solve the relativistic hydrodynamics equations correctly, we then test the capacity of our metric solver.}
Here we will mainly focus on hydrodynamic evolution with \emph{dynamical spacetime}.

\subsection{ \label{sec:BU0} Stability of a stable TOV star}
The first test we perform is the stability of a stable spherically symmetric neutron star BU0.
Even though BU0 is a 1D model, we run this test with a 2-dimensional setup.
The resolution of the simulation is $n_r \times n_{\theta} = 640 \times 64$, where $r = [0,30]$ and $\theta = [0,\pi / 2]$.

\begin{figure}[h]
	\centering
	\includegraphics[width=0.8\columnwidth, angle=0]{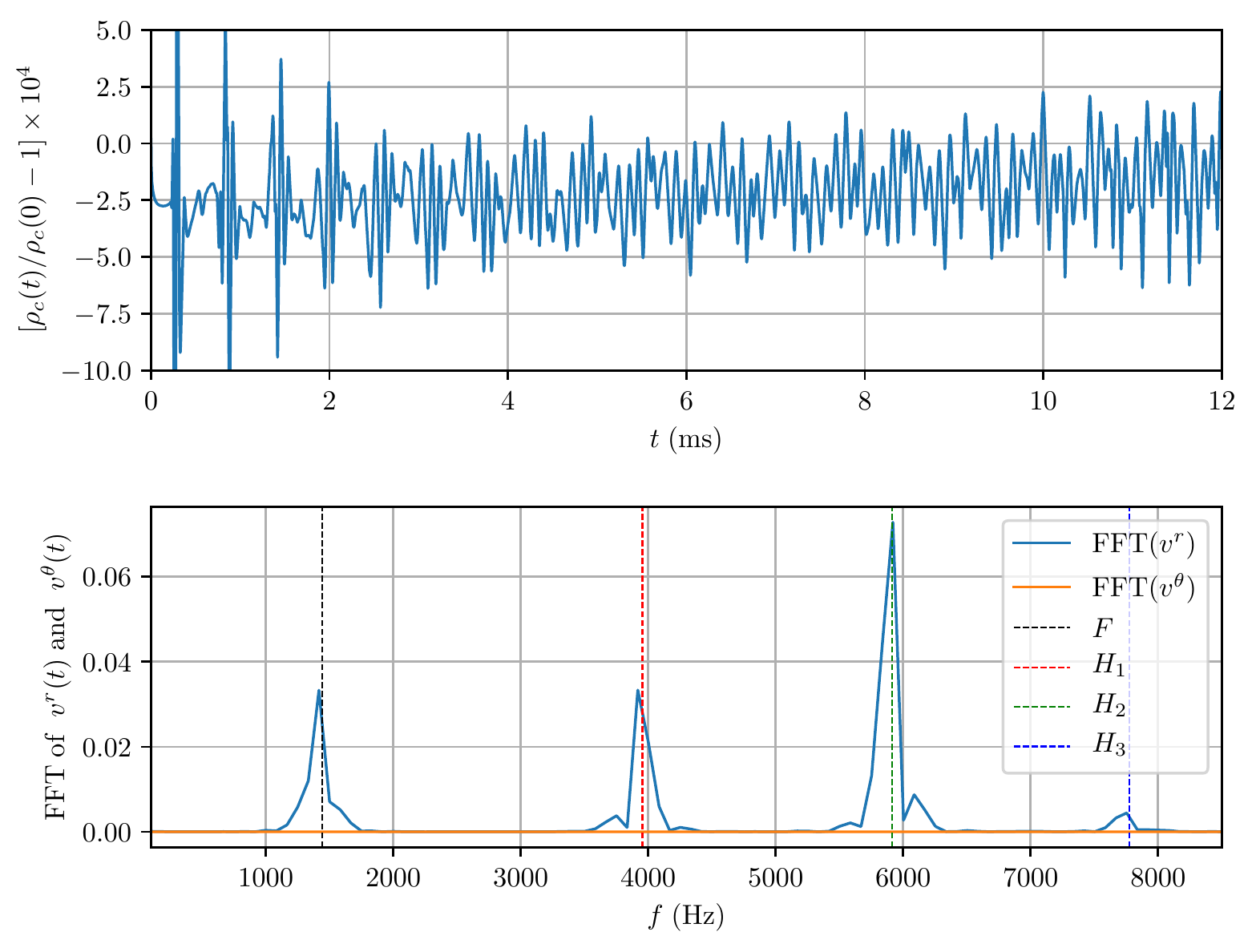}
	\caption{
		A 2-D evolution of a stable spherically symmetric neutron star (TOV star) with the resolution $n_r \times n_{\theta} = 640 \times 64$. 
		\emph{Upper panel}: The relative variations of the centeral density in time.
		The variations are about the order of $O(10^{-4})$.
		\emph{Lower panel}: The Fourier transform of the radial velocity $v^r (t)$ and non-radial velocity $v^{\theta} (t)$ at $r = 5$, $\theta = \pi/4$ (inside the neutron star).
		The vertical lines represent the known eigenmodes frequency.
		Our results agree with the known eigenmodes.
	}
	\label{fig:BU0}	
\end{figure}

While BU0 is a static and stable configuration, the discretization errors and the diffusion at the contact discontinuity of the neutron star surface trigger stellar oscillation modes.
The upper panel of \fref{fig:BU0} shows the relative variation of the central density $\rho_c$ as a function of time.
The relative variation of the density is of the order $10^{-4}$.
The code is able to evolve BU0 stably for more than 10 ms as expected since BU0 is a stable configuration.

One way to test if the code handles a dynamical spacetime correctly, at least in the linear regime for small perturbations, is to extract the eigenmode frequencies from the simulations and compare them with the known values from perturbative calculations.
In order to compare the eigenmodes more clearly, instead of applying the Fourier transform on the central density directly, we analyse the radial component of the velocity $v^r$ and the $\theta$-component $v^{\theta}$ at $r = 5$, $\theta = \pi/4$ (inside the neutron star).
The lower panel of \fref{fig:BU0} shows the Fourier transform of $v^r$ and $v^{\theta}$.
The vertical dashed lines represent the known and well-tested eigenmode frequencies.
Our results agree with the known eigenmode frequencies \cite{2002PhRvD..65h4024F,2006MNRAS.368.1609D, XECHO}.
 {Note that the eigenmode frequencies of an oscillating neutron star in a dynamical spacetime are significantly different from those obtained by Cowling approximation where the metric is kept fixed in time \cite{2006MNRAS.368.1609D}, as shown in \tref{tab:compare_cowling}.}
\begin{table}[h]
  \begin{center}
    \caption{ {The eigenmode frequencies of an oscillating neutron star extracted from our simulations.
The results of the dynamical spacetime case are significantly different from the static spacetime case.} }
    \label{tab:compare_cowling}
    \begin{tabular}{c|c|c|c|c} 
\br
	Dynamical spacetime& $F$(kHz) & $H_1$(kHz) & $H_2$(kHz) & $H_3$(kHz) \\
      \hline
	Yes & $1.417$ & $3.919$ & $5.920$ & $7.753$  \\
	No  & $2.701$ & $4.547$ & $6.303$ & $8.104$  \\
\br
    \end{tabular}
  \end{center}
\end{table}

\subsection{Stability of rotating neutron stars}
The tests we perform in this subsection are the stability of stable rotating neutron stars.
The resolution of these simulation is again $n_r \times n_{\theta} = 640 \times 64$, with $r = [0,30]$ and $\theta = [0,\pi / 2]$.

\begin{figure}[h]
	\centering
	\includegraphics[width=0.8\columnwidth, angle=0]{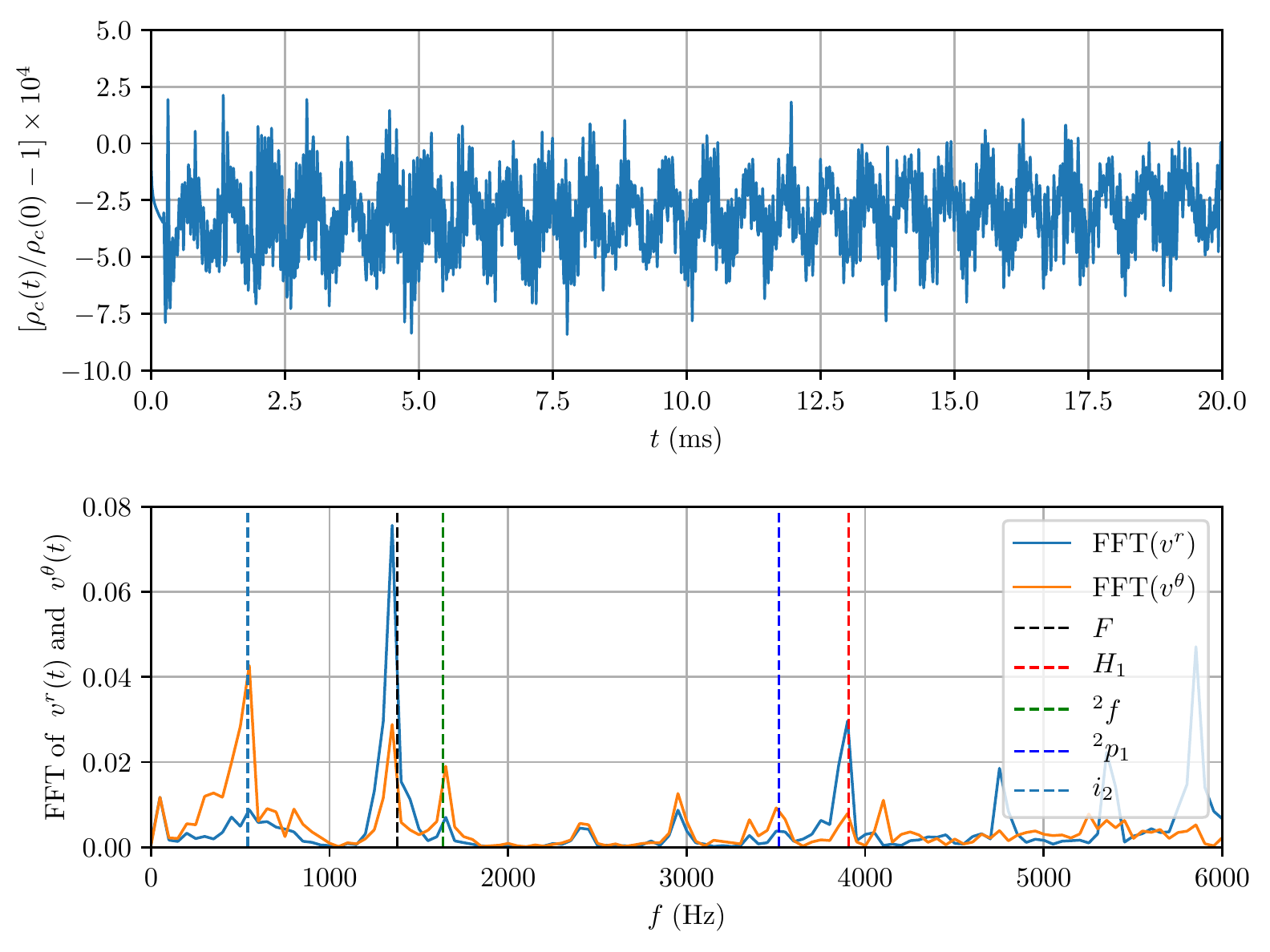}
	\caption{
		The evolution of a stable rotating neutron star (BU2) with the resolution $n_r \times n_{\theta} = 640 \times 64$. 
		\emph{Upper panel}: The relative variation of the centeral density in time.
		The variation is about the order of $O(10^{-4})$.
		\emph{Lower panel}: The Fourier transform of the central density. 
		The vertical lines represent the known and well-tested eigenmode frequencies \cite{2006MNRAS.368.1609D}.
		Our results agree with the known eigenmodes.
	}
	\label{fig:BU2}	
\end{figure}

The upper panel of \fref{fig:BU2} shows the evolution of a rotating stable neutron star BU2.
This model is stably evolved for more than 20 ms, where the relative variation of $\rho_c$ is about the order of $O(10^{-4})$.
The Fourier transforms of $v^r$ and $v^{\theta}$ are shown in the lower panel of the \fref{fig:BU2}.
The extracted eigenmodes again agree with the known results \cite{2006MNRAS.368.1609D}.

\begin{figure}[h]
	\centering
	\begin{subfigure}{0.45\columnwidth}
		\centering
		\includegraphics[width=\columnwidth, angle=0]{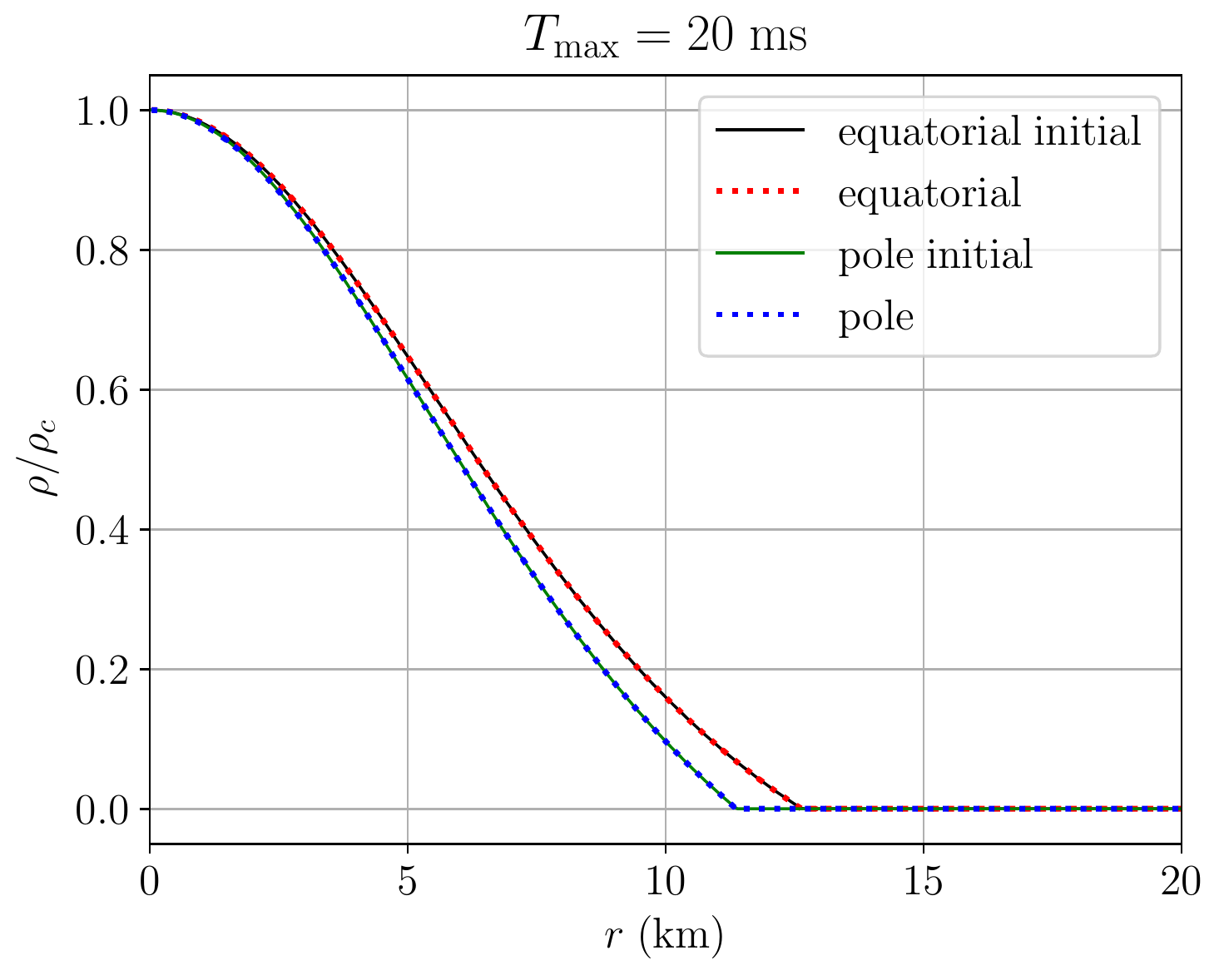}
	\end{subfigure}
	\begin{subfigure}{0.45\columnwidth}
		\centering
		\includegraphics[width=\columnwidth, angle=0]{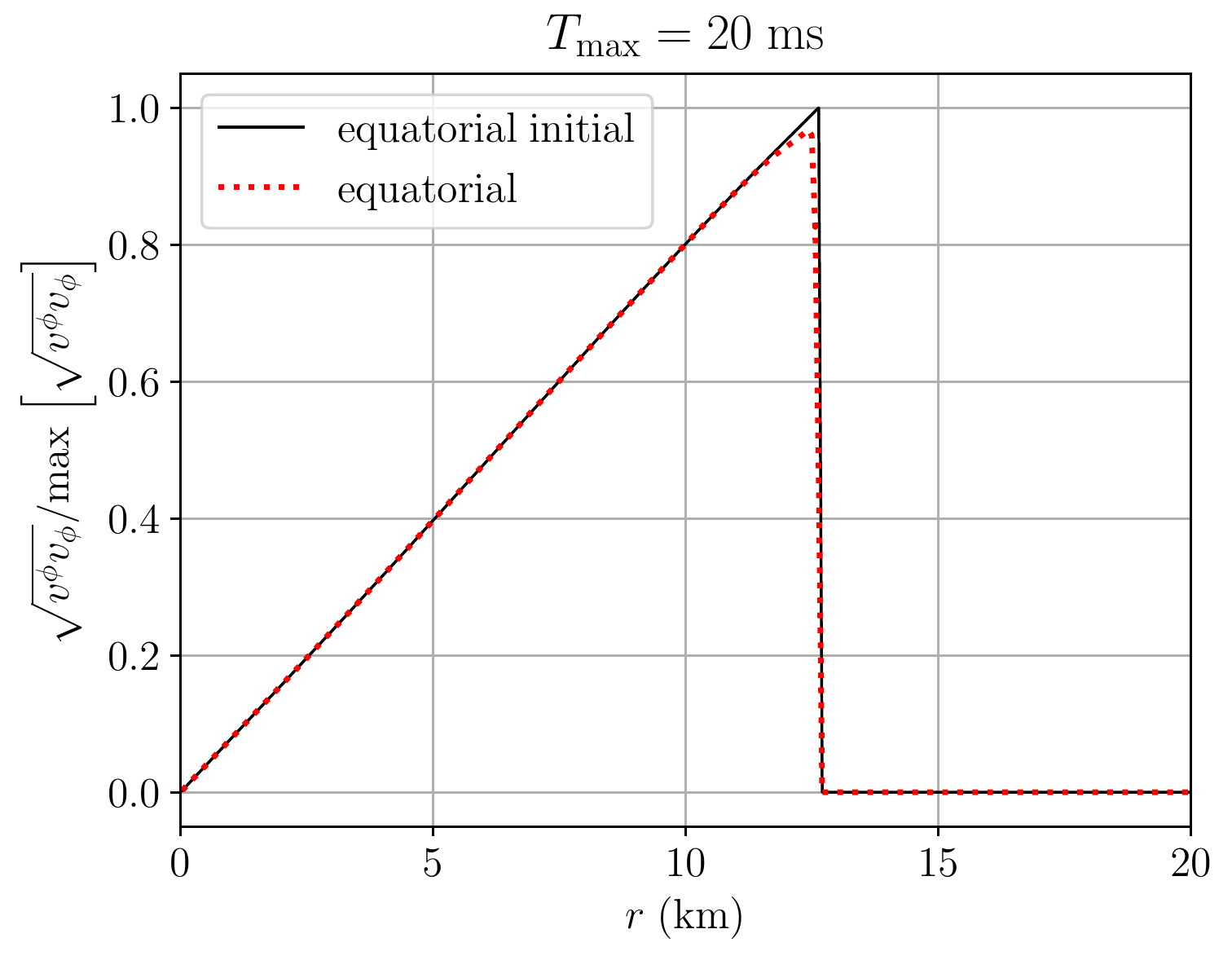}
	\end{subfigure}
	\caption{
		Comparison between the initial density profiles and the rotational velocity (solid lines) and the some quantities (dashed lines) at $T_\text{max} = 20$ ms of a stable rotating neutron star BU2.
	        The left panel shows a comparison between the initial normalized density profiles along the polar and equatorial radii and the same quantities at $T_\text{max}$.
		The right panel compares the rotational velocity profiles at times $t=0$ and $t=T_{\rm max}$.
	}
	\label{fig:BU2_profile}	
\end{figure}

Comparing the density profiles and the rotational velocity profiles at later times with the initial profiles serves as another indicator for code performance.
\Fref{fig:BU2_profile} shows the comparison between the initial density profiles and the rotational velocity (solid lines) with the same quantities (dashed lines) at $T_\text{max} = 20$ ms of BU2.
It can be seen that the density profiles along the equatorial and polar radii are maintained very well during the evolution.
However, the rotational profile is slightly suppressed at the surface of the star.
This is due to the fact that the Riemann solver HLLE we used in these tests is known to be too diffusive to deal with the star surface or any discontinuous surface \cite{XECHO}.

\begin{figure}[h]
	\centering
	\includegraphics[width=0.8\columnwidth, angle=0]{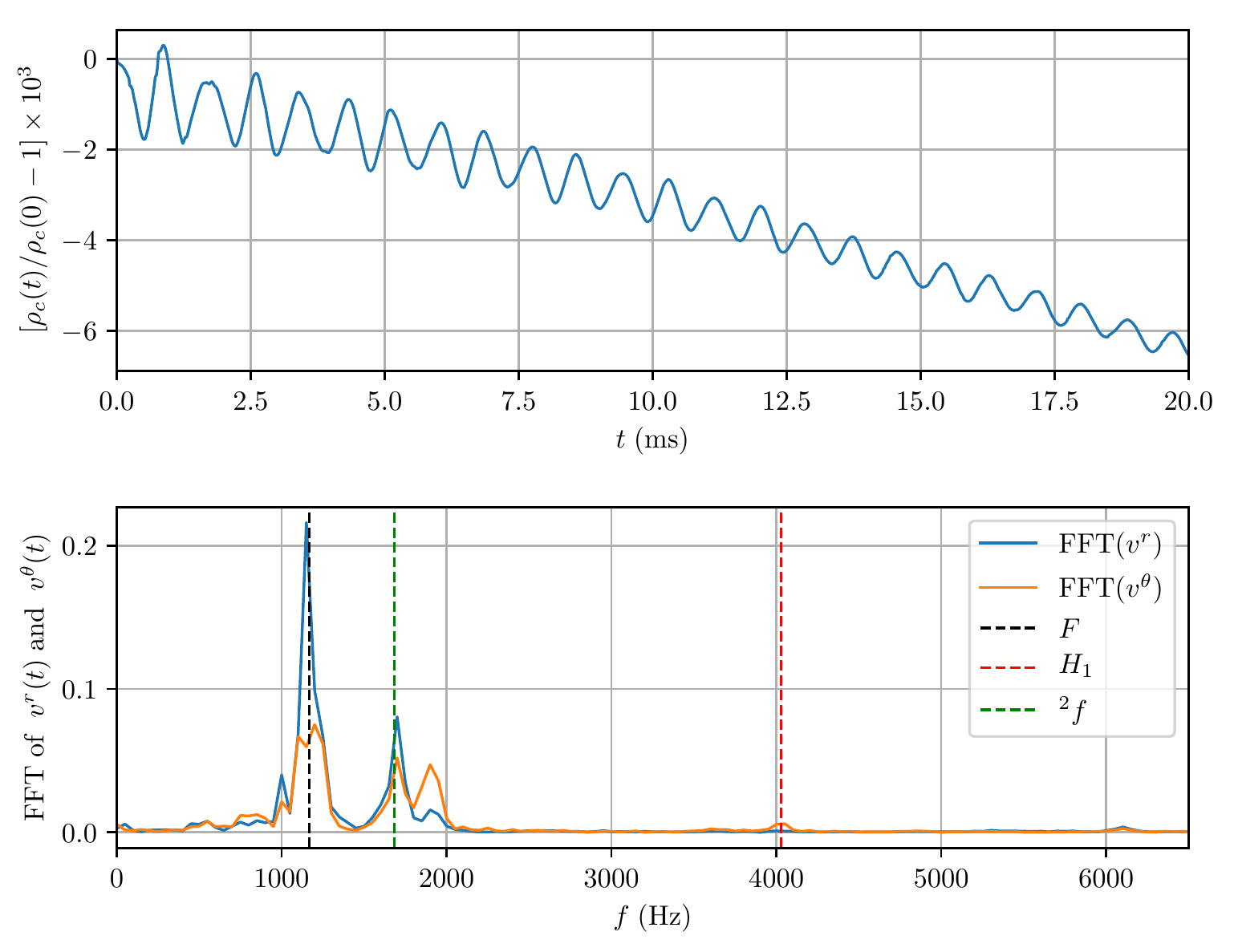}
	\caption{
	        The evolution of a stable rapidly rotating neutron star (BU8) with the resolution $n_r \times n_{\theta} = 640 \times 64$.
                Long term evolution of a rapidly rotating neutron star in a dynamical spacetime is known as a challenging test.
		Even though the HLLE solver might be too diffusive, \texttt{Gmunu} is still able to maintain this model for more than 20 ms with small variations and the fluid behaves correctly and the extracted eigenmodes agree with the results from the other groups \cite{2006MNRAS.368.1609D}.
                \emph{Upper panel}: The relative variation of the central density $\rho_c$ is about {$O(10^{-3})$} and the average value $\rho_c$ {decreases} slowly during the evolution.
		\emph{Lower panel}: The Fourier transform of the radial velocity $v^r (t)$ and non-radial velocity $v^{\theta} (t)$ at $r = 5$, $\theta = \pi/4$ (inside the neutron star).
                The vertical lines represent the known and well-tested eigenmode frequencies \cite{2006MNRAS.368.1609D}.
	}
	\label{fig:BU8}	
\end{figure}

\begin{figure}[h]
	\centering
	\begin{subfigure}{0.45\columnwidth}
		\centering
		\includegraphics[width=\columnwidth, angle=0]{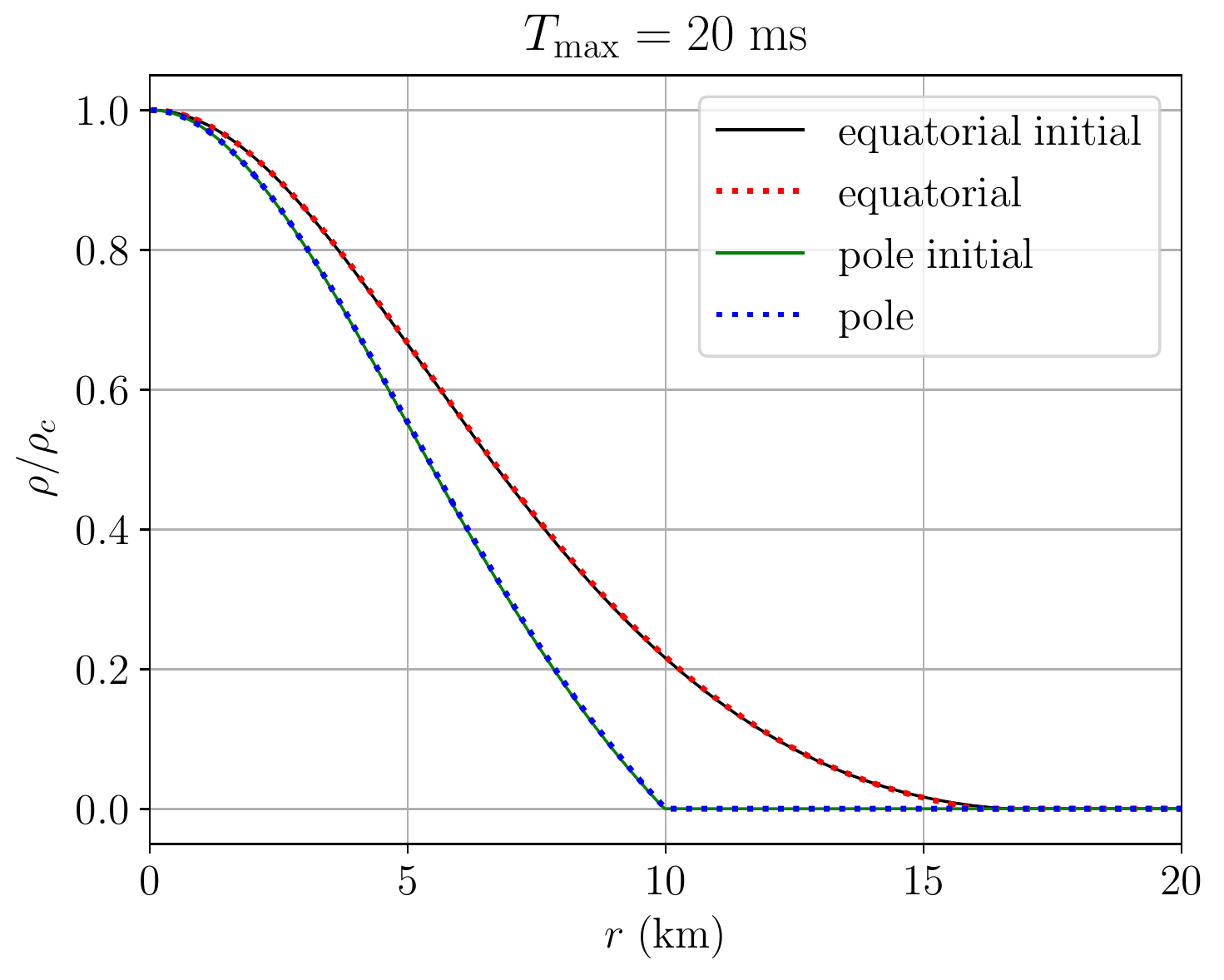}
	\end{subfigure}
	\begin{subfigure}{0.45\columnwidth}
		\centering
		\includegraphics[width=\columnwidth, angle=0]{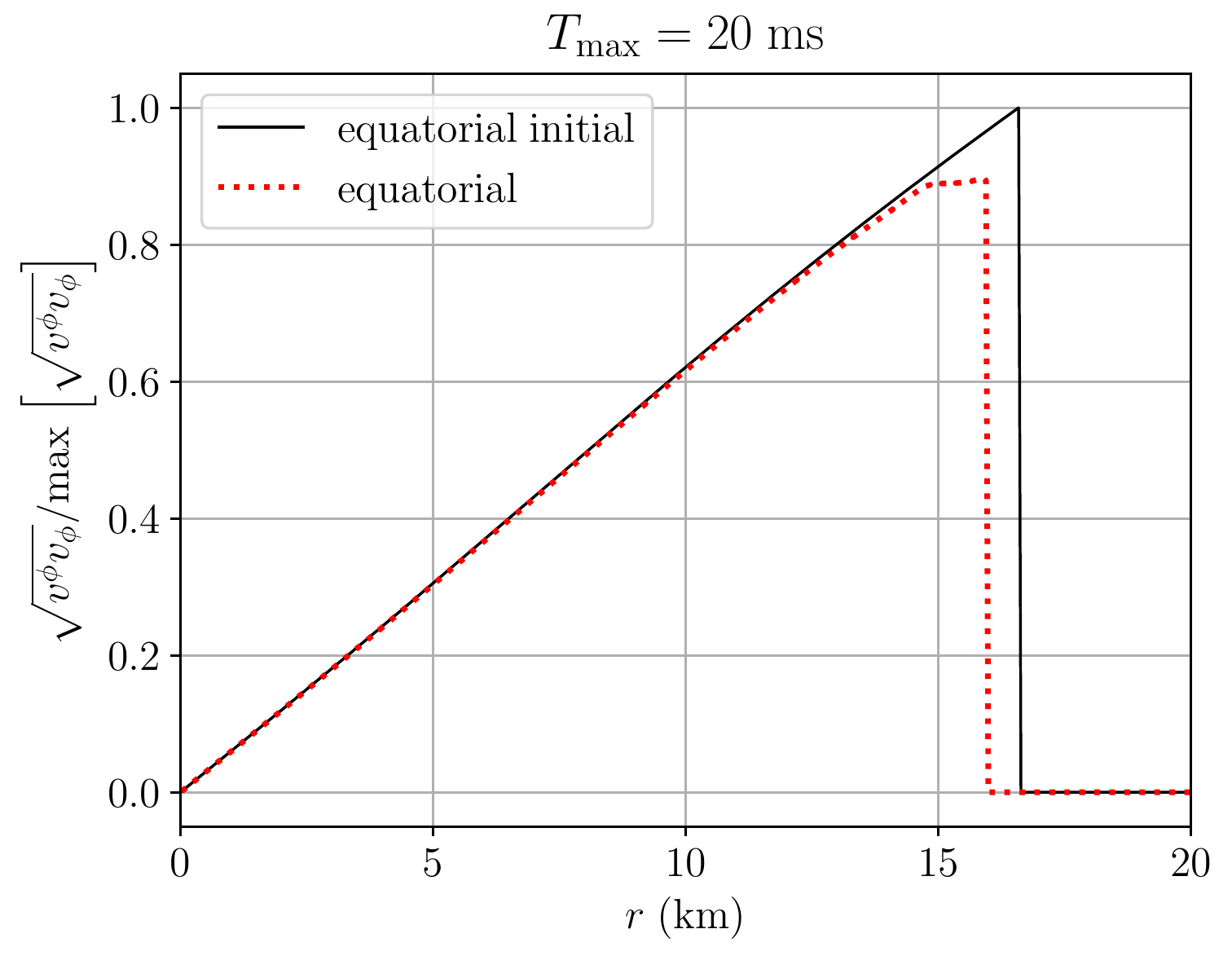}
	\end{subfigure}
	\caption{
                Comparison between the initial density profiles and the normalized rotational velocity (solid lines) and the some quantities (dashed lines) at $T_\text{max} = 20 \ \text{ms}$ of a stable rotating neutron star BU8.
                The left panel shows a comparison between the initial normalized density profiles along the polar and equatorial radii and the same quantities at $T_\text{max}$.
		The right panel compares the rotational velocity profiles at times $t=0$ and $t=T_{\rm max}$.
                Due to the fact that the HLLE solver is too diffusive to deal with the surface of a rapidly rotating star, the angular velocity profile is slightly distorted at the star surface.
	}
	\label{fig:BU8_profile}	
\end{figure}

The same test with the same numerical setup has been done for the model BU8.
Unlike the moderately rotating case BU2, model BU8 is a rapidly rotating neutron star which is close to the mass shedding limit and therefore this test is more demanding.
Even for this demanding case with the diffusive Riemann solver HLLE, \texttt{Gmunu} is able to maintain this model for more than 20 ms.
The relative variation in $\rho_c$ is about {$O(10^{-3})$} as shown in the upper panel of \fref{fig:BU8}.
The oscillation modes extracted from the simulation also agree with the results reported by other groups \cite{2006MNRAS.368.1609D}, as shown in the lower panel in \fref{fig:BU8}.
Moreover, as shown in the left panel in \fref{fig:BU8_profile}, the density profiles are well-preserved even up to $T_\text{max} = 20$ ms.
However, unlike the previous cases, the average value of the central density {decreases} slowly during the evolution, and the angular velocity profile is slightly distorted at the surface of the star.

\subsection{Migration of an unstable TOV star}
Previous tests are all based on the evolution of stable neutron stars, where the configurations are almost stationary with some perturbations.
In this subsection, we report the performance of the code in the fully non-linear regime with significant changes and coupling in the metric and fluid variables.
One of the standard tests for hydrodynamical evolution coupled with dynamical spacetime in the fully non-linear regime is the migration of an unstable neutron star \cite{2002PhRvD..65h4024F, 2010PhRvD..81h4003B, XCFC, XECHO}.
Following \cite{XCFC}, we use model SU for this migration test, which lies on the unstable branch of the mass-radius curve.
In this test, we setup a 1D run with a simulation box $r = [0,30]$ with 1024 grid points.
As the star evolves and migrates to the corresponding stable configuration $\rho_c = 1.346 \times 10^{-3}$ with the same mass, the radius of the star expands to a large value.
Unlike the previous cases, we adopt the ideal gas (gamma-law) equation of state $P = (\Gamma - 1)\rho\epsilon$ with $K = 100$ and $\Gamma = 2$ for the fluid so that we can also capture the shock heating effect.

\begin{figure}[h]
	\centering
	\includegraphics[width=0.8\columnwidth, angle=0]{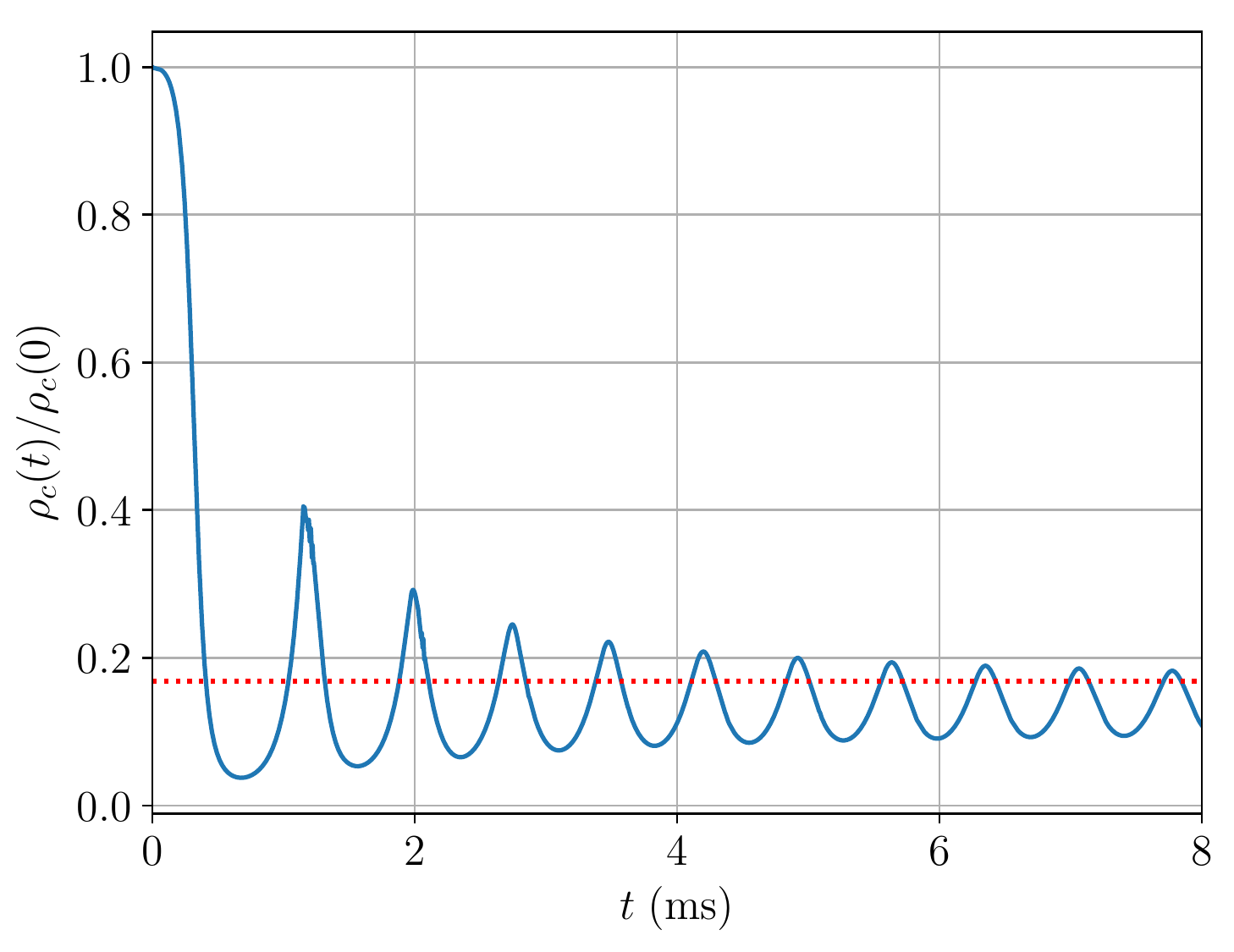}
	\caption{
		Evolution of the central density for an unstable spherically symmetric neutron star with the resolution $n_r \times n_{\theta} = 1024 \times 1$. 
		{The dashed line represents the central density $\rho_c$ of the neutron star on the stable branch.}
	}
	\label{fig:mig}	
\end{figure}

\Fref{fig:mig} shows the evolution of the central density $\rho_c$ as a function of time.
The oscillations are damped due to the fact that shock waves are formed at every pulsation and some kinetic energy is dissipated into thermal energy.
Our results agree with the results from previous works (see, e.g. \cite{XCFC}).

\section{\label{sec:performance}Performance of the metric solver}

\subsection{Convergence properties}
In this section, we demonstrate the convergence properties and performance of our multigrid metric solver.
We solve the metric of model BU8 (see \tref{tab:models}), which represents a rapidly rotating neutron star and far from spherically symmetric, with our metric solver with the resolution $n_r \times n_{\theta} = 640 \times 64$.
The computational domain covers $r = [0,30]$ and $\theta = [0,\pi / 2]$.
To compare the convergence rate, we focus on solving the lapse function $\alpha$ with the flat space initial guess $\alpha = 1$.

\begin{figure}[h]
	\centering
	\includegraphics[width=0.8\columnwidth, angle=0]{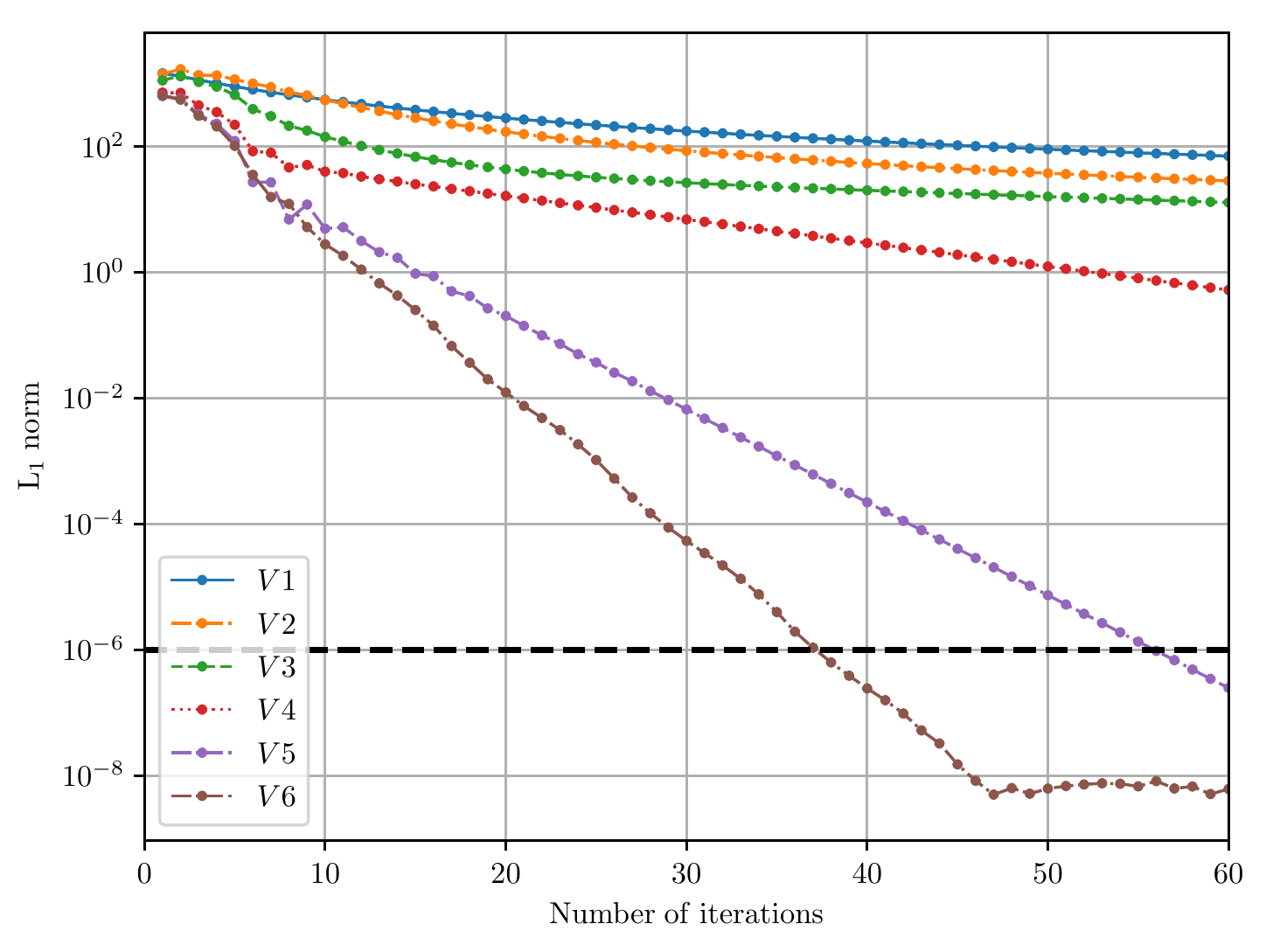}
	\caption{
		$L_1$ norm of residual of Eq.~\eqref{eq:alpha} of an highly non-spherically symmetric model BU8 as a function of the number of iterations with different level of V-cycle.
	The convergence rate increases with the level of the V-cycle.
	Even if the multigrid solver starts from the flat space initial guess, it takes only about 40 iterations for V6 to converge to the prescribed tolerance (horizational black dashed line).
		}
	\label{fig:V-cycles}	
\end{figure}
\Fref{fig:V-cycles} shows the $L_1$ norm of the residual of Eq.~\eqref{eq:alpha} as a function of the number of iterations with different level of V-cycle.
The number represents how deep the solver goes when solving for the lapse function.
The horizontal black dashed line represents the threshold tolerance.
As we can see in \fref{fig:V-cycles}, the solver converges faster when it goes to a deeper level.
Also, it takes $O(10^5)$ iterations (not shown in the plot) to reach the threshold tolerance for the V1 (or Gauss-Seidel) case, while it takes only 37 steps for the V6 case.

In practice, at the beginning of the simulation, we use the initial data provided by \texttt{XNS} as initial guess. During the evolution, we use the previous solution as initial guess for the next iteration.
This makes the solver converge much faster as the solutions on previous time step are usually good approximation to the solution.

\subsection{\label{sec:compare}Frequency of solving the metric}
{Since the metric variables only vary slightly within one hydrodynamics time step in most situations, it is in general not necessary to solve the metric equations at every time step in order to reduce the computational time.
We in practice only solve the metric equations for every few tens of time steps and use extrapolation to obtain the metric quantities in between.
Here we shall study the accuracy and speed of solving the metric with different frequencies.
We introduce the metric resolution parameter $\Delta n$, which represents the number of time steps between solving the metric \cite{CoCoA}.
To see how $\Delta n$ affects the accuracy and the performance of our code, we performed simulations with the same setting \tref{tab:setting} but with different metric resolution parameter, i.e. $\Delta n = 5, 10, 30, 50$.}
\begin{table}[h]
  \begin{center}
    \caption{
	  {The simulation setting we used for metric resolution parameter analysis.}
	  }
    \label{tab:setting}
    \begin{tabular}{ll} 
\br
	Model & BU8 \\
	Riemann solver & HLLE \\
	reconstruction scheme & MC \\
	$n_r \times n_{\theta}$ & $640 \times 64$ \\
	range of $r$ & $[0,30]$ \\
	range of $\theta$ &$[0,\pi / 2]$ \\
	$T_{\rm{max}}$ $[\rm{ms}]$ & 20\\
\br
    \end{tabular}
  \end{center}
\end{table}

\begin{figure}[h]
	\centering
	\includegraphics[width=0.8\columnwidth, angle=0]{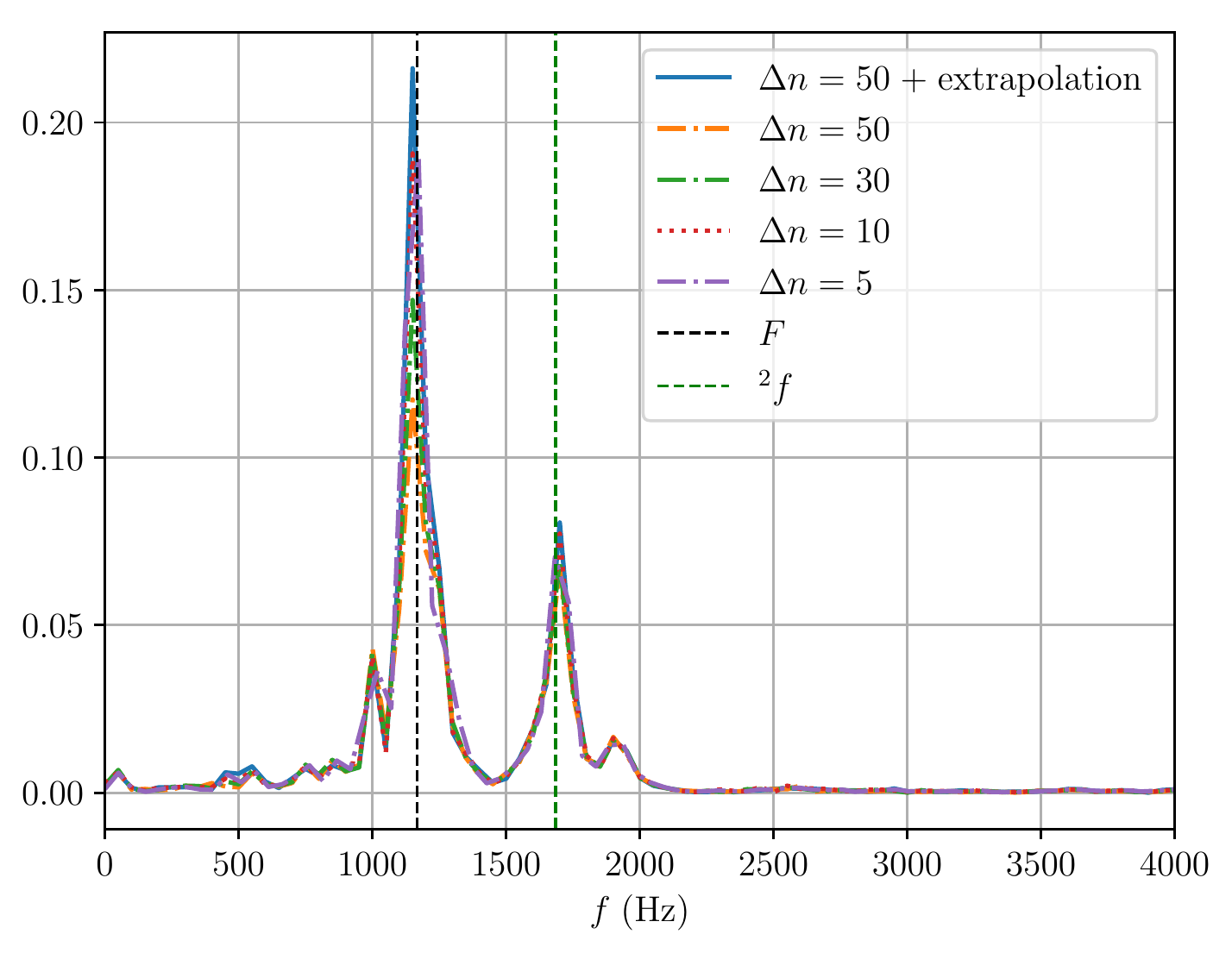}
	\caption{
                {The Fourier transform of the radial velocity $v^r (t)$ at $r = 5$, $\theta = \pi/4$ (inside the neutron star BU8) of different $\Delta n$.
		For more detailed simulation setting, see \tref{tab:setting}.
        The vertical dashed lines represent the known eigenmode frequencies.
        All the extracted eigenmodes frequencies agree with the known results even for $\Delta n = 50$ without metric extrapolation case.
		However, although the results have no significant differences between these settings, the computational cost is significantly high if the metric resolution parameter $\Delta n$ is too small (see \Tref{tab:profiling}).}
	}
	\label{fig:n_steps_compare}	
\end{figure}

{\Fref{fig:n_steps_compare} shows the Fourier transform of the radial velocity $v^r (t)$ at $r = 5$, $\theta = \pi/4$ (inside the neutron star) of different metric resolution parameter $\Delta n$.
The vertical dashed lines represent the known eigenmode frequencies.
As shown in \fref{fig:n_steps_compare}, the eigenmode frequencies are almost the same for all $\Delta n$.
This result demonstrates that it is not necessary to set a small $\Delta n$ for mildly dynamical spacetimes such as oscillating rotating neutron stars.
We leave the study of more generic dynamical spacetimes for future investigation.
Since the computational cost of extrapolation is much smaller than the metric solver (see below), it is employed in our code to obtain the metric quantities for times within each "step" $\Delta n$.
Here we note that 4-points Lagrange interpolation is used for metric extrapolation in \texttt{Gmunu}. }

{ The computational cost of our metric solver is comparable to the hydrodynamics solver.
To demonstrate this, we again perform the same simulations but the maximum time step $n$ is limited to $10^4$, with code profiler.
\Tref{tab:profiling} shows the percentages of time spent on different routines on each hydrodynamic step.
As expected, the results show that with smaller $\Delta n$, the time spent on the multigrid (MG) metric solver is longer.
For the case $\Delta n = 50$ and extrapolating the metric in between, the computational cost of our metric solver is roughly 1.22 times larger than the hydrodynamics solver.
It can also be seen from the table that the computational time is dominated by the metric solver if $\Delta n \sim O(1)$.
}
\begin{table}[h]
  \begin{center}
    \caption{
	{The percentages of time spent on different routines on each hydrodynamic step.
	These are the simulations of BU8 with HLLE and MC limiter where the maximum time step $n$ is limited to $10^4$.
	The ``hydro step'' here includes only the hydrodynamics solver and the multigrid (MG) metric solver, other routines such as data output is not included in this step.
	Note that we ignored the routines of which the contribution is less than $1\%$. }
	  }
    \label{tab:profiling}
    \begin{tabular}{|c|l|c|c|c|c|} 
	\hline
	\multicolumn{2}{|c|}{$\Delta n$} & \textbf{50} & \textbf{30} & \textbf{10} & \textbf{5}\\
	\hline
	\multirow{4}*{ $r_{\rm{hydro}}$ [\%] } & Riemann solver & 11.21 & 8.77 & 4.11 & 2.61 \\
	 & reconstruction & 19.69 & 15.4 & 7.74 & 4.59 \\
	 & source terms & 8.40 & 6.57 & 3.30 & 1.96 \\
	 & con2prim & 4.30 & 3.35 & 1.65 & $\leq$ 1.00 \\
	\hline
	\multirow{2}*{ $r_{\rm{metric}}$ [\%] } & MG solver & 48.87 & 60.08 & 80.11 & 88.40\\
	 & extrapolation & 6.02 & 4.66 & 2.19 & 1.15 \\
	\hline
	\multicolumn{2}{|c|}{$r_{\rm{metric}} / (1 - r_{\rm{metric}})$} & \textbf{1.22} & \textbf{1.84} & \textbf{4.65} & \textbf{8.57}\\
	\hline
    \end{tabular}
  \end{center}
\end{table}

{Note that these tests were performed with the most basic (also the fastest) available setting of hydrodynamics solver (HLLE and MC limiter) on a rapidly rotating neutron star BU8.
While in practical use, we often use more accurate hydrodynamics solver and reconstruction scheme such as WENO5 or MP5 which make the corresponding computational cost for the hydrodynamics higher.
Moreover, unlike the rapidly rotating model BU8, the spacetimes of most of the isolated neutron stars deviate not too much from spherical symmetry.
In these cases, the multigrid solver usually converges faster than the highly-asymmetric cases.
In other words, for modelling isolated neutron stars, we expect that the computational cost of our metric solver is almost the same as (sometimes even lower than) the hydrodynamics solver.}

\section{\label{sec:conclusion}Conclusion}
   
We present the methodology and implementation of \texttt{Gmunu}, a new general-relativistic hydrodynamics code which makes use of cell-centered nonlinear multigrid methods to solve the elliptic-type metric equations in the extended conformally flat condition (xCFC) approximation to general relativity. 
The set of hydrodynamics equations are solved with standard high-resolution shock-capturing schemes.
Four different Riemann solvers have been implemented in the code: TVDLF \cite{1996ApL&C..34..245T}, HLL \cite{Harten1997}, HLLE \cite{HLLE1, HLLE2}, and Marquina flux formula \cite{1996JCoPh.125...42D}. 
For the cell-interface reconstruction, various options are also available: PC, MC, WENO5, and MP5.

We have tested \texttt{Gmunu} with some benchmarking tests for relativistic hydrodynamics codes such as the relativistic shocktube problem, the evolution of rapidly rotating neutron stars, and the migration from an unstable TOV star to a corresponding stable solution with the same mass.

The main novelties of \texttt{Gmunu} are the following:

\begin{itemize}
        \item Although the system is highly nonlinear and fully coupled, our multigrid solver is robust and converges rapidly.
                In practical use {(e.g. solving the metric in every 50 steps, and extrapolating in between)}, the computational time needed for solving the (elliptic-type) metric equations is comparable to the hydrodynamics step. 
        \item In contrast to the code presented in \cite{CoCoNuT} where the hydrodynamic and metric variables are defined in two different grids, our multigrid metric solver uses the same grid as the hydrodynamics sector.
		This avoids the need to perform interpolations for the variables defined in two different grids.
        \item Even using spherical polar coordinates, besides standard boundary conditions, our code does not require special treatment or regularization near the origin and pole axis. 
\end{itemize}

We have demonstrated that multigrid method is an efficient strategy to solve nonlinear elliptic metric equations in hydrodynamical simulations. 
It seems to be promising that the method would make fully-constrained evolution scheme in numerical relativity become more affordable computationally. 

In the future, we shall extend \texttt{Gmunu} to a fully-constrained scheme in exact general relativity such as the formulation of Bonazzola {\it et al.} \cite{2004PhRvD..70j4007B}. 
While numerical-relativity codes based on a free-evolution approach are standard choices for modelling dynamical spacetimes, it is still challenging for these codes to preform stable and accurate long-term evolutions of compact objects. 
It would be interesting to see whether a fully-constrained evolution code could improve the situation without requiring much more computational resources in the future.

{Not only this work can be further investigated on the numerical relativity side, \texttt{Gmunu} can also be entended for realistic astrophysical applications such as core-collapse supernova simulations.
Future work includes code parallelisation, the development of full 3D metric solver and microphysics implementation. }

\section{Acknowledgements}
PCKC thanks Elias Most for the useful discussion about the details of conserved to primitive variables, and Ninoy Rahman for the detailed discussion about the special treatment at the centre of the simulation box.
This work was partially supported by grants from the Research Grants Council of the Hong Kong (Project No. CUHK 14310816 and CUHK 24304317), the Croucher Innovation Award from the Croucher Fundation Hong Kong and by the Direct Grant for Research from the Research Committee of the Chinese University of Hong Kong.

\appendix 

\section{\label{appendix:metric}Implementation of the metric equations}
By following Ref.~\cite{XCFC,XECHO}, Eqs.~\eqref{eq:X}-\eqref{eq:beta} can be solved in the given order.
In the flat spacetime, the Laplacian of a scalar function $u(r,\theta)$ is 
\begin{equation}
	\Delta u = \frac{\partial^2 u}{\partial r^2} + \frac{2}{r}\frac{\partial u}{\partial r} + \frac{1}{r^2}\left(\frac{\partial^2 u}{\partial \theta^2} + \cot\theta \frac{\partial u}{\partial \theta}\right).
\end{equation}
In our implementation for the vector equations, we solve for the orthonormal-basis components for vector fields instead of coordinate-basis components.   
We rewrite a generic vector as
\begin{align}
	&X^{\hat{r}} := X^{r} ,\\
	&X^{\hat{\theta}} := rX^{\theta} ,\\
	&X^{\hat{\phi}} := r\sin\theta X^{\phi}.
\end{align}
The conformal vector Laplacian (the left hand side of Eqs.~\eqref{eq:X} and \eqref{eq:beta}) are
\begin{align}
	&(\Delta \bm{X})^{\hat{r}} = \Delta X^{\hat{r}} - \frac{2}{r^2}\left(X^{\hat{r}} + \frac{\partial X^{\hat{\theta}}}{\partial \theta} + \cot\theta X^{\hat{\theta}}\right) + \frac{1}{3}\frac{\partial }{\partial r} \left( \nabla_j  X^j \right),  \\
	&{(\Delta \bm{X})^{\hat{\theta}} = \Delta X^{\hat{\theta}} + \frac{2}{r^2}\frac{\partial X^{\hat{r}}}{\partial \theta} - \frac{X^{\hat{\theta}}}{r^2\sin^2 \theta} + \frac{1}{3r}\frac{\partial }{\partial \theta} \left( \nabla_j  X^j \right), }\\
	&{(\Delta \bm{X})^{\hat{\phi}} = \Delta X^{\hat{\phi}} - \frac{X^{\hat{\phi}}}{r^2\sin^2 \theta}, }
\end{align}
where the divergence of the vector $\bm{X}$ is
\begin{equation}
	\nabla_j X^j = \frac{\partial X^{\hat{r}}}{\partial r} + \frac{1}{r} \left( 2X^{\hat{r}} + \frac{\partial X^{\hat{\theta}}}{\partial \theta} + \cot\theta X^{\hat{\theta}}\right).
\end{equation}

Note that from the numerical point of view, Eqs.~\eqref{eq:X} and \eqref{eq:beta} are the same equations with different source terms.
On the other hand, it is better to solve the scalar equations, Eqs.~\eqref{eq:psi} and \eqref{eq:alpha}, for the deviation of the functions from their asymptotic-flatness limits
$\psi \rightarrow 1$ and $\alpha \rightarrow 1$ due to the non-linearity of the equations.
For instance, instead of solving for the conformal factor $\psi$ directly, we solve for its deviation $\delta \psi := \psi - 1$. 

\subsection{Boundary conditions at the origin and the axis}
We first discuss the inner boundary ($r \rightarrow 0$) and also the boundary conditions at the axis ($\theta = 0$ or $\theta = \pi/2$).
For the scalar variables, as they have to be continuous across all boundaries, we impose the symmetric boundary condition for all boundaries except at the outer boundary ($r = r_{\text{max}}$).
However, the boundary conditions for vectors are non-trival.
Here we followed the approach in Ref.~\cite{CoCoA}, which is summerized in \tref{tab:boundary}.
\begin{table}[h]
  \begin{center}
    \caption{ Boundary conditions for the vector components at centre and axis. 
	  The plus sign ``$+$" means symmetric boundary condition whereas the minus sign ``$-$" represent anti-symmetric boundary condition.}
    \label{tab:boundary}
    \begin{tabular}{l|c|c|c} 
\br
	& $\beta^r$ & $\beta^\theta$ & $\beta^\phi$ \\
      \hline
	centre & $-$ & $+$ & $+$  \\
	pole & $+$ & $-$ & $+$  \\
	equator & $+$ & $-$ & $+$ \\
\br
    \end{tabular}
  \end{center}
\end{table}

\subsection{{Discretization}}
{In \texttt{Gmunu}, the metric solver are formulated in second-order accuracy.
In particular, the finite differences of the derivatives in \texttt{Gmunu} are the following:
\begin{align}
	\frac{\partial u}{ \partial r}\Big|_{i,j} &= \frac{u_{i+1,j} - u_{i-1,j}}{2 \Delta r},\\
	\frac{\partial u}{ \partial \theta}\Big|_{i,j} &= \frac{u_{i,j+1} - u_{i,j-1}}{2 \Delta \theta},\\
	\frac{\partial^2 u}{ \partial r^2}\Big|_{i,j} &= \frac{u_{i+1,j} -2u_{i,j} + u_{i-1,j}}{ \Delta r^2},\\
	\frac{\partial^2 u}{ \partial \theta^2}\Big|_{i,j} &= \frac{u_{i,j+1} - 2u_{i,j} + u_{i,j-1}}{ \Delta \theta^2},\\
	\frac{\partial^2 u}{ \partial \theta \partial r}\Big|_{i,j} &= \frac{u_{i+1,j+1} - u_{i+1,j-1}- u_{i-1,j+1} + u_{i-1,j-1}}{ 4 \Delta \theta \Delta r}.
\end{align}
}

\section*{References}
\bibliographystyle{abbrv}
\bibliography{cite.bib}

\end{document}